\documentclass[amssymb,aps,twocolumn,floats,showpacs]{revtex4-1}
\usepackage{color,graphicx,pstricks,float}
\usepackage{natbib}

\usepackage{fancybox}
\usepackage{epsfig}
\usepackage{graphicx}
\usepackage{tabularx}
\usepackage{inputenc}
\usepackage{amsmath}
\usepackage{amssymb}
\usepackage{gensymb}
\usepackage{array}

\usepackage{color,graphicx,pstricks,float}

\begin{document}

\title{Rashba-Zener mechanism for nanoscale skyrmions and topological metals}

\author{Deepak S. Kathyat}
\author{Arnob Mukherjee}
\author{Sanjeev Kumar}

\address{ Department of Physical Sciences,
Indian Institute of Science Education and Research Mohali, Sector 81, S.A.S. Nagar, Manauli PO 140306, India \\
}

\begin{abstract}
We report a microscopic electronic mechanism for nanoscale skyrmion formation and topological metalicity. The mechanism, which relies on combining 
the classic double-exchange (DE) physics with the Rashba spin orbit coupling (SOC), not only provides an accurate understanding of existence of skyrmions but also explains key features in small angle neutron scattering (SANS) and Lorentz transmission electron microscopy (LTEM) data on thin films of a variety of magnetic metals. 
The skyrmion states are characterized as disordered topological metals via explicit calculations of Bott index and Hall conductivity. 
Local density of states (LDOS) display characteristic oscillations that are shown to be arising from a combination of confinement effect and gauge-field induced Landau level physics. The presence of oscillations in LDOS, without external magnetic flux,
is a direct consequence of the Rashba-Zener (RZ) mechanism. 
The results are based on hybrid simulations on a model that explicitly retains itinerant electronic degrees of freedom. A simple physical picture is provided via an effective short-range spin model with coupling constants that depend on electronic kinetic energy.
The mechanism reported here not only opens up a new approach to understand skyrmion formation in metals, but also 
provides a guiding principle for discovering exotic topological metal states.
\end{abstract}
\date{\today}

\maketitle

\section*{Introduction}
Magnetic skyrmions are being envisioned as building blocks of next-generation data storage and processing devices \cite{Fert2017, Wiesendanger2016, Fert2013, Nagaosa2013b, Gobel2019, Bogdanov2020}.
This possibility has led to a surge in research activity geared towards identifying candidate materials \cite{Dupe2014, Pollard2017, Soumyanarayanan2017, Romming2013b, Yu2012, Yu2011, Zhao2016, Meyer2019, Tonomura2012, Hirschberger2019,Jin2017,Karube2016, Muhlbauer2009b}. Such textures in metals are particularly important since they can be manipulated by ultra-low electrical currents \cite{Song2020, Sampaio2013, Romming2013b, Yu2012}. Appearance of sparse as well as packed skyrmions has been reported in thin films of a variety of chiral metallic magnets \cite{Nayak2017, Yu2011, Zhao2016, Pfleiderer2004, Jena2020, Meyer2019, Hsu2018, Tonomura2012, Yu2018, Nagase2019}.
However, the current understanding of skyrmion formation in magnets is via spin Hamiltonians that either include Dzyaloshinskii-Moriya (DM) interactions or geometrical frustration
\cite{Rossler2006, Chen2016, Mohanta2019, Zang2011a, Iwasaki2014}. This approach is inconsistent for metals as the aforementioned spin Hamiltonians usually originate from a Mott insulating state.
Therefore, the importance of electronic Hamiltonian based understanding of skyrmion formation in metals has been recognized and a mechanism based on RKKY interactions has recently been put forward \cite{Ozawa2017a, Wang2020}. 

Experimentally, skyrmions are typically stabilized in thin film magnets upon application of external field perpendicular to the surface. Many experimental studies show two peculiar features 
at magnetic field values lower than those required for skyrmion formation -- a diffuse ring pattern in small angle neutron scattering (SANS) experiments and filamentary domain walls in Lorentz transmission electron microscopy (LTEM) experiments \cite{Pollard2017, Soumyanarayanan2017, Romming2013b, Yu2012, Yu2011, Zhao2016, Meyer2019, Tonomura2012,Hirschberger2019,Jin2017,Karube2016, Yu2018, Nagase2019}. These features seem to be clear precursors for skyrmion formation, and the corresponding phase may be viewed as the parent state of skyrmions.
At present, a microscopic explanation of these experimental features does not exist. Most DM interaction based theories indicate that a spin spiral state with ordering wave vector ($Q,Q$) is the parent of the skyrmion state. 
Introduction of DE mechanism by Zener represents a milestone in our understanding of ferromagnetic metals \cite{Zener1951a, Anderson1955, deGennes1960a}. The mechanism has played a key role in the description of magnetic and magneto-transport phenomena across families of materials, such as, perovskite manganites, dilute magnetic semiconductors and Heusler metals \cite{Dagotto2002, Pradhan2017, Yaouanc2020, Bombor2013}. 
Surprisingly, the role of DE physics in skyrmion formation has largely remained unexplored.

In this work, we show that the DE mechanism combined with the Rashba SOC provides an accurate microscopic understanding of existence of skyrmions in magnetic metals with large local moments. We explicitly demonstrate, via the state-of-the-art hybrid Monte Carlo (HMC) simulations, the appearance of skyrmions in the Rashba DE (RDE) model. An effective spin Hamiltonian is studied via large scale Monte Carlo simulations for a comprehensive understanding of the origin of these spin textures. We find magnetic states hosting sparse skyrmions (sSk) as well as packed skyrmions (pSk) of Neel type, in addition to an ordered skyrmion crystal phase. A filamentary domain wall (fDW) phase is identified as the parent of sSk, and a single-Q (SQ) spiral state leads to pSk. These findings are consistent with SANS and LTEM data on thin films of Co-Zn-Mn alloys, FeGe and MnSi, and transition metal multilayers \cite{Romming2013b, Yu2012, Yu2011, Zhao2016, Karube2016, Yu2018, Nagase2019}. Furthermore, we find that the skyrmion phases are natural realizations of topological metals as characterized by explicit calculations of the Bott index and the topological Hall conductivity. We identify features in LDOS that are unique to the proposed mechanism, hence providing clear testable predictions for the presence of RZ mechanism in real systems.

\section*{Results}
\subsection*{Skyrmions in Rashba double-exchange model}
Starting with the ferromagnetic Kondo lattice model (FKLM) in the presence of Rashba SOC on a square lattice and taking the double-exchange limit, we obtain the RDE Hamiltonian \cite{Kathyat2020a},

\begin{eqnarray}
H_{\rm RDE} &=& \sum_{\langle ij \rangle, \gamma} [g^{\gamma}_{ij} d^{\dagger}_{i} d^{}_{j} + {\rm H.c.}] - h_z \sum_i S^z_i,
\label{Ham-RDE}
\end{eqnarray}
\noindent
where, $d^{}_{i} (d^{\dagger}_{i})$  annihilates (creates) an electron at site ${i}$ with spin parallel to the localized spin. The second term represents the Zeeman coupling of local moments to external magnetic field of strength $h_z$. Site $j = i + \gamma$ is the nn of site $i$ along spatial direction $\gamma = x,y$. The projected hopping $g^{\gamma}_{ij}$ depend on the orientations of the local moments ${\bf S}_i$ and ${\bf S}_j$. The tight-binding, $t^{\gamma}_{ij}$ and Rashba, $\lambda^{\gamma}_{{ij}}$, contributions to $g^{\gamma}_{ij} = t^{\gamma}_{ij} + \lambda^{\gamma}_{ij}$ are given by \cite{Kathyat2020a},

\begin{eqnarray}
t^{\gamma}_{ij} & = & -t \big[\cos(\frac{\theta_i}{2}) \cos(\frac{\theta_j}{2}) 
+ \sin(\frac{\theta_i}{2})  \sin(\frac{\theta_j}{2})e^{-\textrm{i} (\phi_i-\phi_j)} \big],
\nonumber \\
\lambda_{{ij}}^{x} & = & \lambda \big[\sin(\frac {\theta_i}{2})  \cos(\frac {\theta_j}{2})e^{-\textrm{i} \phi_i} - \cos(\frac {\theta_i}{2})  \sin(\frac {\theta_j}{2})e^{\textrm{i} \phi_j}\big],
\nonumber \\ 
\lambda_{{ij}}^y & = & \textrm{i} \lambda \big[\sin(\frac {\theta_i}{2})  \cos(\frac {\theta_j}{2})e^{-\textrm{i} \phi_i} + \cos(\frac {\theta_i}{2})  \sin(\frac {\theta_j}{2})e^{\textrm{i} \phi_j}\big],
\end{eqnarray}

\noindent
where $\theta_i$ ($\phi_i$) is the polar (azimuthal) angle for localized moment ${\bf S}_i$. The strengths of hopping $t$, and Rashba SOC $\lambda$ are parametrized by $\alpha$ as $t = (1-\alpha) t_0$ and $\lambda = \alpha t_0$, where $t_0=1$ sets the reference energy scale.

We study the RDE Hamiltonian using the state-of-the-art hybrid Monte Carlo (HMC) simulations  (see "Methods"). Since our main focus is to search for skyrmions in $H_{{\textrm RDE}}$, we present results at low temperatures with increasing Zeeman field using zero field cooled (ZFC) protocol. Presence of skyrmions is inferred via local skyrmion density \cite{Chen2016},
\begin{eqnarray} 
\chi_{i} & = & \frac{1}{8\pi} [ {\bf S}_i \cdot ({\bf S}_{i+x} \times {\bf S}_{i+y} ) + {\bf S}_i \cdot ({\bf S}_{i-x} \times {\bf S}_{i-y})],
\end{eqnarray}
\noindent
which is the discretized version of the continuum definition, $ {\bf S} \cdot (\partial_{x}{\bf S} \times \partial_{y}{\bf S}) /4\pi$. Total skyrmion density is defined as, $\chi = \sum_i \chi_i$.  
We also compute the spin structure factor (SSF),
\begin{eqnarray}
S_{f}({\bf q}) &=& \frac{1}{N^2}\sum_{ij} {\bf S}_i \cdot {\bf S}_j~ e^{-{\rm i}{\bf q} \cdot ({\bf r}_i - {\bf r}_j)},
\label{SSF}
\end{eqnarray}
\noindent
and the relevant component of vector chirality $\eta$ as,
\begin{eqnarray} 
\eta & = & \frac{1}{N} \sum_{i} ({\bf S}_{i} \times {\bf S}_{i+x} )\cdot \hat{y} - ({\bf S}_{i} \times {\bf S}_{i+y} )\cdot \hat{x}.
\end{eqnarray} 
\noindent
Averaging of all quantities over MC steps is implicitly assumed, unless stated otherwise.

\begin{figure}
\includegraphics[width=.96 \columnwidth,angle=0,clip=true]{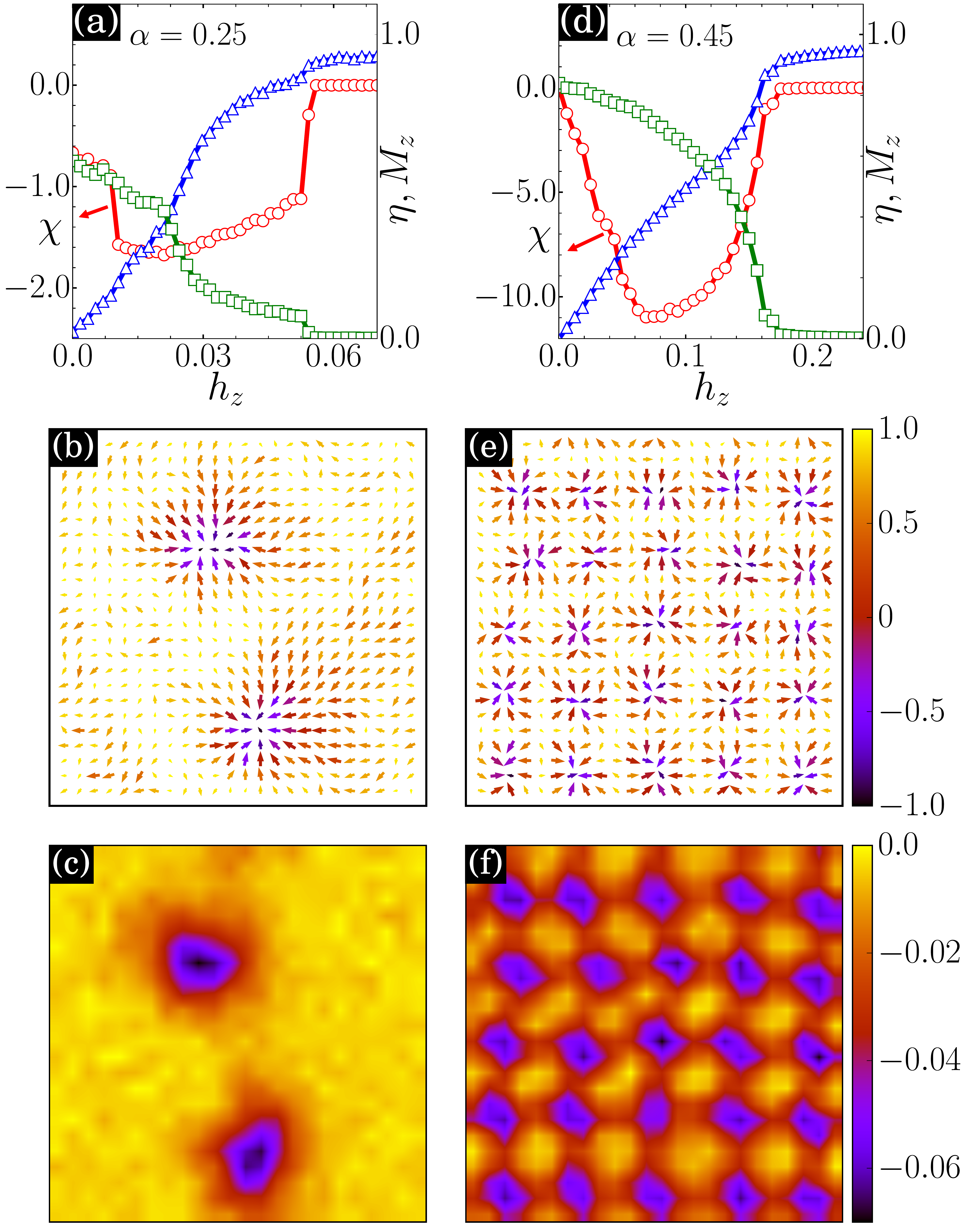}
\caption{{\bf Skyrmions in the Rashba double-exchange Hamiltonian.}
Magnetization $M_z$ (triangles), total skyrmion density $\chi$ (circles) and vector chirality $\eta$ (squares) as a function of applied Zeeman field for, (a) $\alpha = 0.25$, and (d) $\alpha = 0.45$. Snapshots of spin configurations, (b), (e), and the local skyrmion density, (c), (f), at $T=0.01$ for representative values of $\alpha$ and $h_z$: (b)-(c) $\alpha =0.25 $, $h_z=0.03$; (e)-(f) $\alpha=0.45$, $h_z=0.09$. The color bar for spin configurations represents the $S_z$ value. 
}
\label{fig1}
\end{figure}

Results obtained via HMC simulations for two representative values of $\alpha$ are shown in Fig. \ref{fig1}. 
Magnetization, $M_z = \frac{1}{N} \sum_i S^z_i$, increases upon increasing $h_z$, as expected. $\eta$ starts out with a finite positive value at $h_z=0$,
decreases monotonically upon increasing $h_z$, and finally vanishes in the saturated ferromagnet (sFM) state. The magnitude of $\chi$ initially increases with applied field, and then decreases on approach to the sFM state (see circles in Fig. \ref{fig1}(a), (d)). 
The qualitative behaviour appears to be similar between $\alpha = 0.25$ and $\alpha=0.45$.
The negative sign of $\chi$ reveals that the polarity of skyrmions is opposite to the orientation of the background magnetization. 

The existence of skyrmion states in the electronic model Eq. (\ref{Ham-RDE}) is explicitly demonstrated via the spin configurations as well as skyrmion density maps. We find that small values of $\alpha$ lead to sparse skyrmions (see Fig. \ref{fig1}(b)), and the packing (size) of skyrmions increases (decreases) with increasing $\alpha$ (see Fig. \ref{fig1}(e)). The negative polarity is consistent with the fact that the central spin in the skyrmion texture is oriented opposite to the magnetization direction (see Fig. \ref{fig1}(c), (f)). We also note that the skyrmions obtained here are of Neel type with negative effective magnetic monopole charge. 

Having demonstrated via state of the art computations that the RDE Hamiltonian hosts sparse and packed skyrmions, we now focus on understanding the origin of these textures. While HMC is a very powerful method for explicit simulations of electronic Hamiltonians, by itself it does not provide a simple understanding of the results.
Therefore, we study an effective spin model derived from the RDE Hamiltonian and identify distinct conditions for the formation of sparse and packed skyrmions.

\subsection*{Sparse and packed skyrmions in the effective microscopic spin Hamiltonian}
Including the Zeeman coupling term in the recently derived effective spin model for $H_{{\textrm RDE}}$ \cite{Kathyat2020a, Banerjee2014}, we obtain,

\begin{eqnarray}   \label{eq:ESH} 
H_{\rm eff} & = &-\sum_{\langle ij \rangle, \gamma} D^{\gamma}_{ij} f^{\gamma}_{ij} - h_z \sum_i S^z_i,  \nonumber \\
\sqrt{2} f^{\gamma}_{ij} & = &  \big[ t^2(1+{\bf S}_i \cdot {\bf S}_j) + 2t\lambda  \hat{\gamma'} \cdot ({\bf S}_i \times{\bf S}_j)  \nonumber  \\
& & + \lambda^2(1-{\bf S}_i \cdot {\bf S}_j+2 (\hat{\gamma'} \cdot {\bf S}_i)(\hat{\gamma'} \cdot {\bf S}_j)) \big]^{1/2}, \nonumber \\
D^{\gamma}_{ij} & = & \langle [e^{{\rm i} h^{\gamma}_{ij}} d^{\dagger}_{i} d^{}_{j} + {\textrm H.c.}] \rangle_{gs}.
\end{eqnarray} 

\noindent
In the above, $\hat{\gamma'} = \hat{z} \times \hat{\gamma}$, $f^{\gamma}_{ij}$ ($h^{\gamma}_{ij}$) is the modulus (argument) of complex number $g^{\gamma}_{ij}$ and $\langle \hat{O} \rangle_{gs}$ denotes expectation values of operator $\hat{O}$ in the ground state. 
It has been shown that using a constant value of $D^{\gamma}_{ij}$ captures the essential physics of the Hamiltonian Eq. (\ref{eq:ESH}), therefore we set $D^{\gamma}_{ij} \equiv D_0 = 1$ in our simulations \cite{Kathyat2020a}.

\begin{figure}
\includegraphics[width=.96 \columnwidth,angle=0,clip=true]{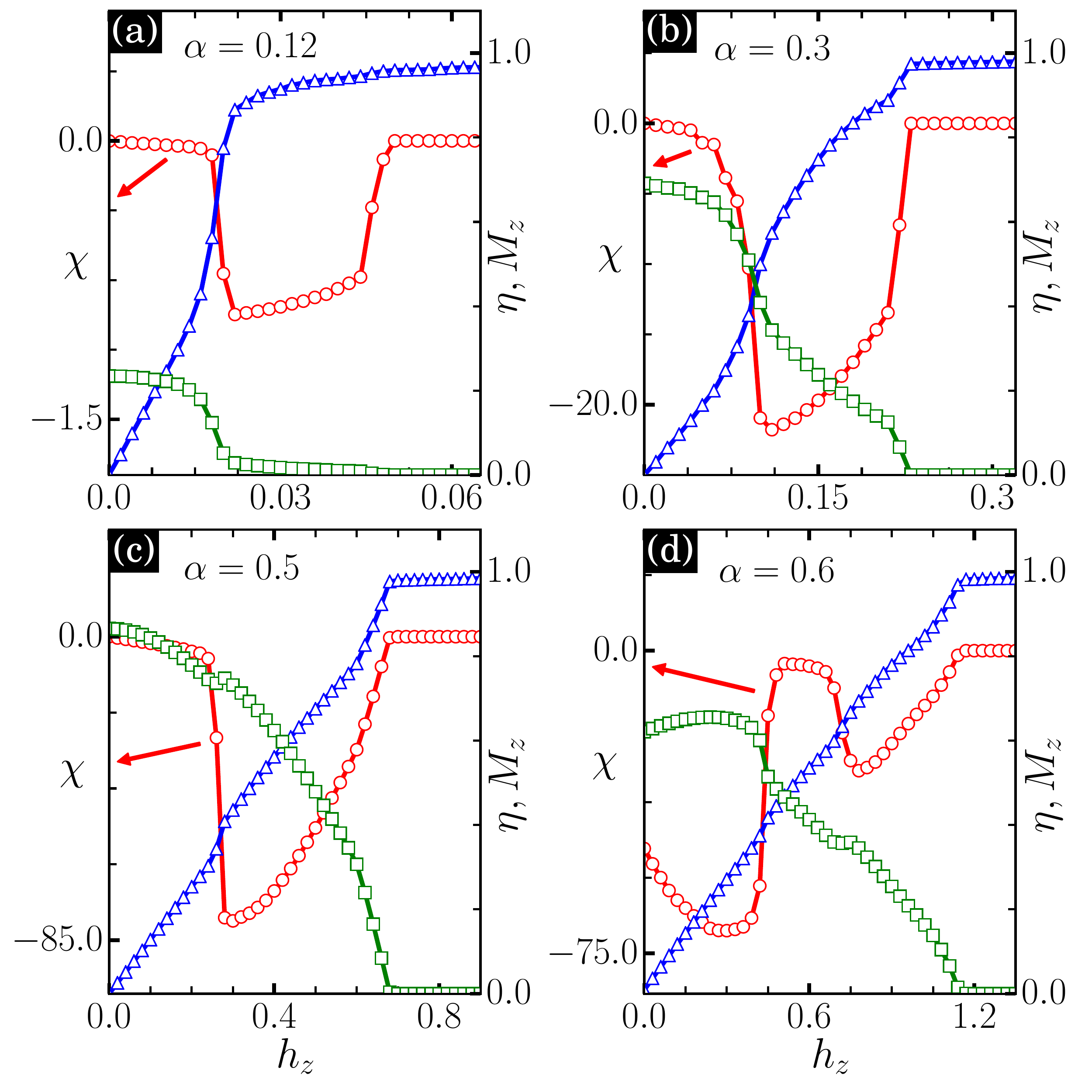}
\caption{{\bf Bulk characterization via effective spin Hamiltonian.} (a) - (d) Magnetization (triangles), total skyrmion density (circles) and vector chirality (squares) as a function of applied Zeeman field for different values of $\alpha$. Left $y$-axis scale is for $\chi$.
}
\label{fig2}
\end{figure}

A direct test for the validity of the effective spin model is to check if $H_{\rm eff}$ also supports skyrmion formation with increasing Zeeman field.
We simulate $H_{\rm eff}$ using the standard classical MC scheme (see "Methods"). We find that the field-dependence of magnetization, $\eta$ and $\chi$ for $H_{\rm eff}$ is similar to that obtained via HMC (compare Fig. \ref{fig1} (a), (d) and Fig. \ref{fig2}). 
\begin{figure}
\includegraphics[width=.96 \columnwidth,angle=0,clip=true]{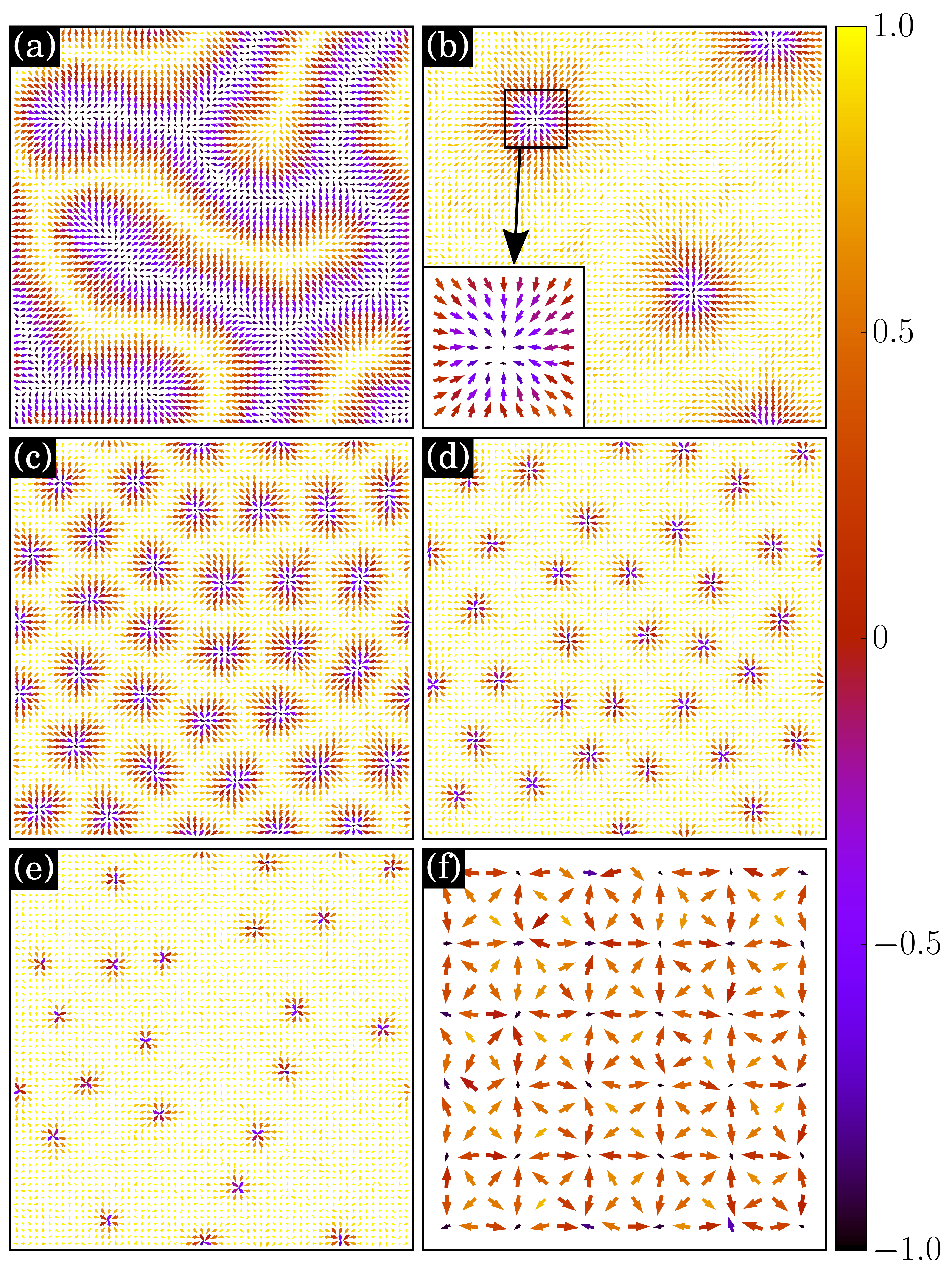}
\caption{{\bf Sparse and packed skyrmions.} 
Low temperature snapshots of spin configurations for values of $\alpha$ and $h_z$ representative of different phases: (a) fDW state at $\alpha=0.16$, $h_z=0$; (b) sparse skyrmions at $\alpha=0.16$, $h_z=0.036$; (c) pSk at $\alpha=0.32$, $h_z=0.13$ ; (d) pSk at $\alpha=0.32$, $h_z=0.21$;
(e) sSk at $\alpha=0.32$, $h_z=0.25$; (f) square SkX at $\alpha=0.6$, $h_z=0.3$. For clarity, only $16 \times 16$ section of the $60 \times 60$ lattice is displayed in (f).
}
\label{fig3}
\end{figure}
\begin{figure}
\includegraphics[width=.96 \columnwidth,angle=0,clip=true]{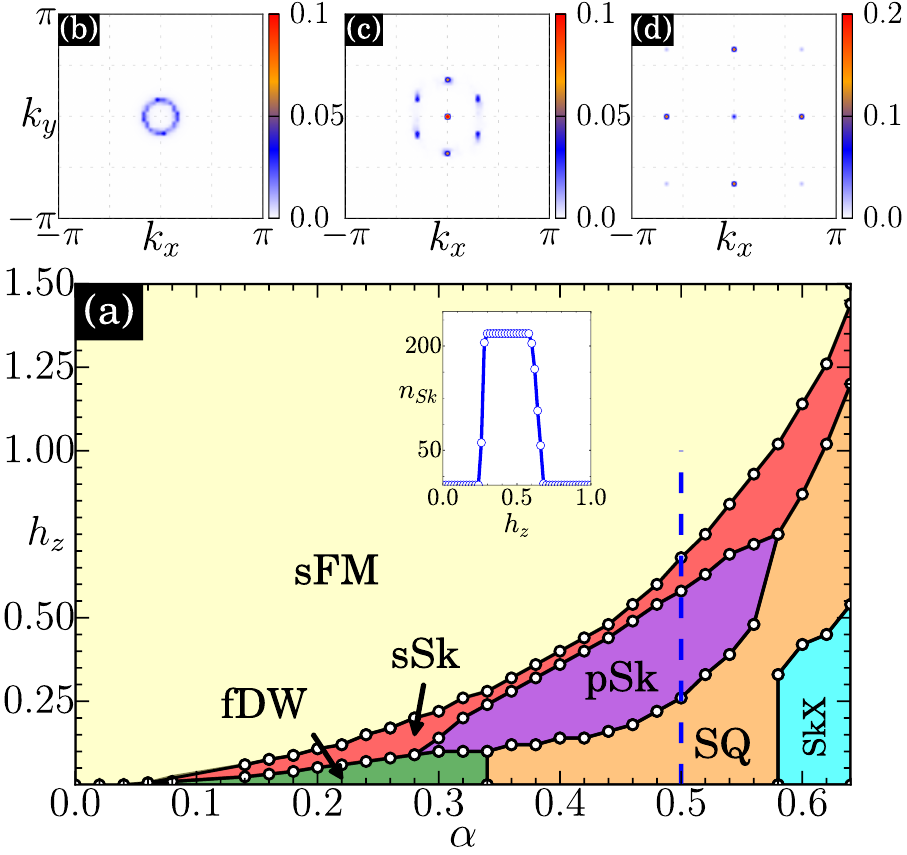}
\caption{{\bf Phase diagram.} 
(a) Ground state phase diagram in the $\alpha$-$h_z$ plane. SSF for, (b) fDW at $\alpha = 0.22$, $h_z = 0$ (c) pSk at $\alpha = 0.4$, $h_z = 0.16$, and (d) SkX at $\alpha = 0.6$, $h_z = 0.3$.
Inset in (a) shows an explicit count of skyrmion centers, $n_{\textrm{Sk}}$, as a function of $h_z$ along the vertical dashed line at $\alpha=0.5$. Plateau in $n_{\textrm{Sk}}$ coincides with the pSk phase. 
}
\label{fig4}
\end{figure}
For small values of $\alpha$, magnetization increases linearly for small $h_z$, followed by a slower than linear rise. This change to non-linear behaviour is accompanied by a sharp increase in the magnitude of $\chi$ (see Fig. \ref{fig2}(a), (b)). A simple understanding is that the emergence of skyrmions arrests the ease with which spins align along the direction of external magnetic field. 
A finite value of $\eta$ in the absence of magnetic field originates from the DM-like terms present in our effective Hamiltonian.
Variation of $\eta$ is anticorrelated with that of magnetization and the former shows a sharp decrease accompanying the increase in magnitude of $\chi$ (see Fig. \ref{fig2}(a), (b)). Finally, for still larger values of applied field, system approaches sFM state, with both $\chi$ and $\eta$ vanishing. For $\alpha = 0.5$, the change in $\chi$ near $h_z = 0.25$ is sharper, and is accompanied by a weak discontinuity in both magnetization and $\eta$ (see Fig. \ref{fig2}(c)). This qualitatively different behaviour is an indicator of the pSk state, as will be illustrated below with the help of real space spin configurations.
For $\alpha = 0.6$, $\chi$ is finite even at $h_z = 0$. This is consistent with our results reported for Rashba FKLM \cite{Kathyat2020a}. Interestingly, the magnitude of $\chi$ reduces with increasing $h_z$, and then again increases before finally vanishing on approach to the sFM state (see Fig. \ref{fig2}(d)). The re-entrant behaviour of $\chi$ suggests that the SkX state reported here cannot simply be viewed as an ordered arrangement of pre-formed skyrmions. 

We show in Fig. \ref{fig3} the evolution of magnetic textures with change in $\alpha$ and $h_z$ within $H_{\rm eff}$. 
We find fDW states in the absence of external field for small $\alpha$ (see Fig. \ref{fig3}(a)) \cite{Kathyat2020a}. We observe that the junctions of these domains
turn out to be natural nucleation centers for skyrmions when magnetic field is applied (see Supplementary Information). 
For small values of $\alpha$, the skyrmions are sparse (see Fig. \ref{fig3}(b)), and $\chi$ increases with $\alpha$ leading to pSk phase (see Supplementary Information). For a given $\alpha$, increasing $h_z$ leads, initially, to a reduction of the size by polarizing the spins in the peripheral region of skyrmions (compare Fig. \ref{fig3} (c) and (d)) and then to a reduction of the number (compare Fig. \ref{fig3} (d) and (e)). A perfectly ordered crystal of smallest possible skyrmions on a square lattice is obtained in the absence of external field at $\alpha = 0.6$ (see Fig. \ref{fig3}(f)). We have also confirmed that the skyrmion formation in the model is not an artifact of the ZFC protocol, by verifying their existence using the field cooled protocol (see Supplementary Information).

We now summarize the results discussed above in the form of a phase diagram in Fig. \ref{fig4}(a). 
We identify the following qualitatively distinct regimes, in addition to the trivial sFM state: (i) a fDW state, (ii) a state with sSk, (iii) a SQ spiral with peaks in the spin structure factor at $(0,Q)$ or $(Q,0)$, (iv) a pSk state, and (v) a SkX with square geometry. The boundaries separating these regimes are inferred from variations in $\chi$, $\eta$ and magnetization, as described in Fig. \ref{fig2}. The SSF for fDW, pSk and SkX states are displayed in Fig. \ref{fig4}(b)-(d), in that order. 
Circular diffuse pattern for small $\alpha$ (see Fig. \ref{fig4}(b)-(c)) matches well with SANS experiments and Fourier transform of LTEM images on MnSi and Co-Zn-Mn alloys \cite{Yu2012, Tonomura2012, Karube2016}. We also characterize the pSk state by plotting the number of skyrmions, $n_{Sk}$, obtained by explicitly counting skyrmion centers, as a function of applied field (see inset in Fig. \ref{fig4}(a)). The constancy of $n_{Sk}$ is an indicator of the pSk state.

\subsection*{Topological metalicity and confined states}
Finally, we discuss some unique topological features of the skyrmion states obtained via RZ mechanism. While insulating topological states in translationally invariant systems have been theoretically very well studied, the possibility of finding topological metallic or insulating phases in disordered systems has been proposed only recently \cite{Agarwala2017,Yang2019}. The proposed models, however, are not easy to realize as they involve an unusual dependence of hopping amplitudes on the relative orientation of lattice vectors. We find that the skyrmion phases of the RDE model are direct realizations of topological metallic states. 
\begin{figure*}
\includegraphics[width=1.88 \columnwidth,angle=0,clip=true]{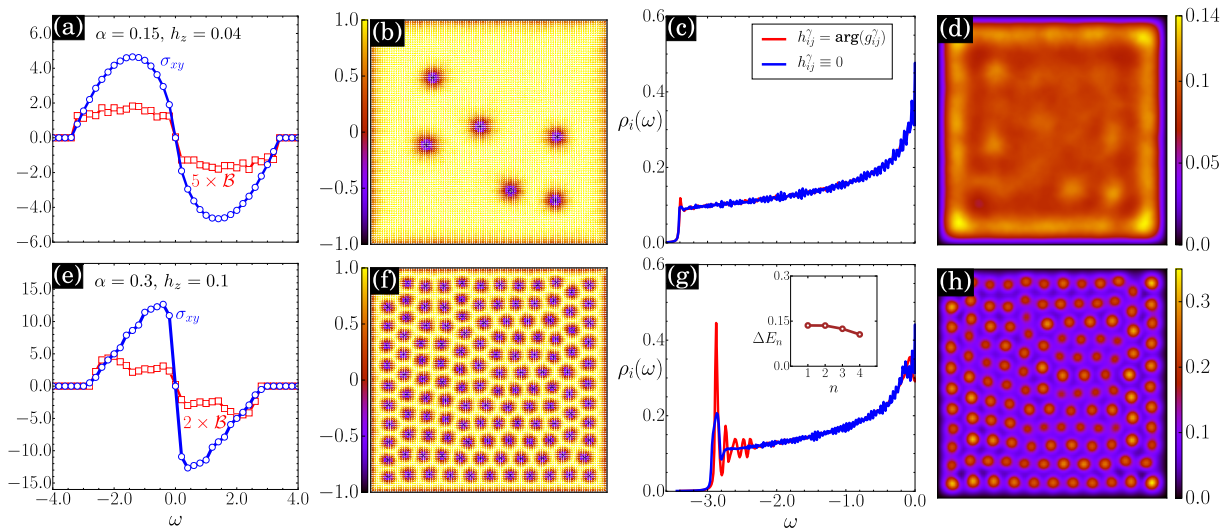}
\caption{{\bf Topological Hall effect and local density of states.} 
(a) Bott index, ${\cal B}$, and Hall conductivity, $\sigma_{xy}$, in units of $e^2/h$ across the band, (b) low temperature magnetic configuration obtained via simulations with open boundary conditions, (c) local density of states in skyrmion cores with (red lines) and without (blue lines) local gauge fields, and (d) real space map of LDOS at $\omega = -3.38$ in absence of local gauge fields ($h^{\gamma}_{ij} \equiv 0$). Panels (a)-(d) display results for $\alpha = 0.15$. (e)-(h) Same quantities as shown in (a)-(d), in that order, for $\alpha = 0.30$. LDOS in panel (h) is shown for $\omega = -2.87$.
}
\label{fig5}
\end{figure*}
We show this by explicitly providing a topological characterization of the sparse and packed skyrmion states by computing the Bott index ${\cal B}$ (see "Methods"). We also compute Hall conductivity $\sigma_{xy}$ using the standard Kubo-Greenwood formula (see "Methods"). Both sSk and pSk states support finite values of $\sigma_{xy}$ as well as ${\cal B}$ (see Fig. 5(a), (d)). Moreover, a clear correlation between $\sigma_{xy}$ and ${\cal B}$ confirms that the Hall effect present in the skyrmion phases is of purely topological origin. Indeed, it should be noted that the external magnetic field, coupled to localized spins via Zeeman term, is important only for stabilizing the skyrmion states and does not contribute to Hall effect in our calculations. While the normal and topological Hall contributions are typically mixed in experiments, they can be separated by observing the field and temperature dependence of the Hall response \cite{Ritz2013}.
We take the analogy with Hall systems one step further by investigating the effect of boundary conditions on the skyrmion states. Using open boundary conditions in simulations, we find that skyrmions in the bulk remain intact while the textures on the boundary are drastically modified (see Fig. 5(b), (f)). The lattice boundary seems to display incomplete skyrmion textures, reminiscent of incomplete cyclotron orbits along the edges of quantum Hall systems.

\begin{figure*}
\includegraphics[width=1.88 \columnwidth,angle=0,clip=true]{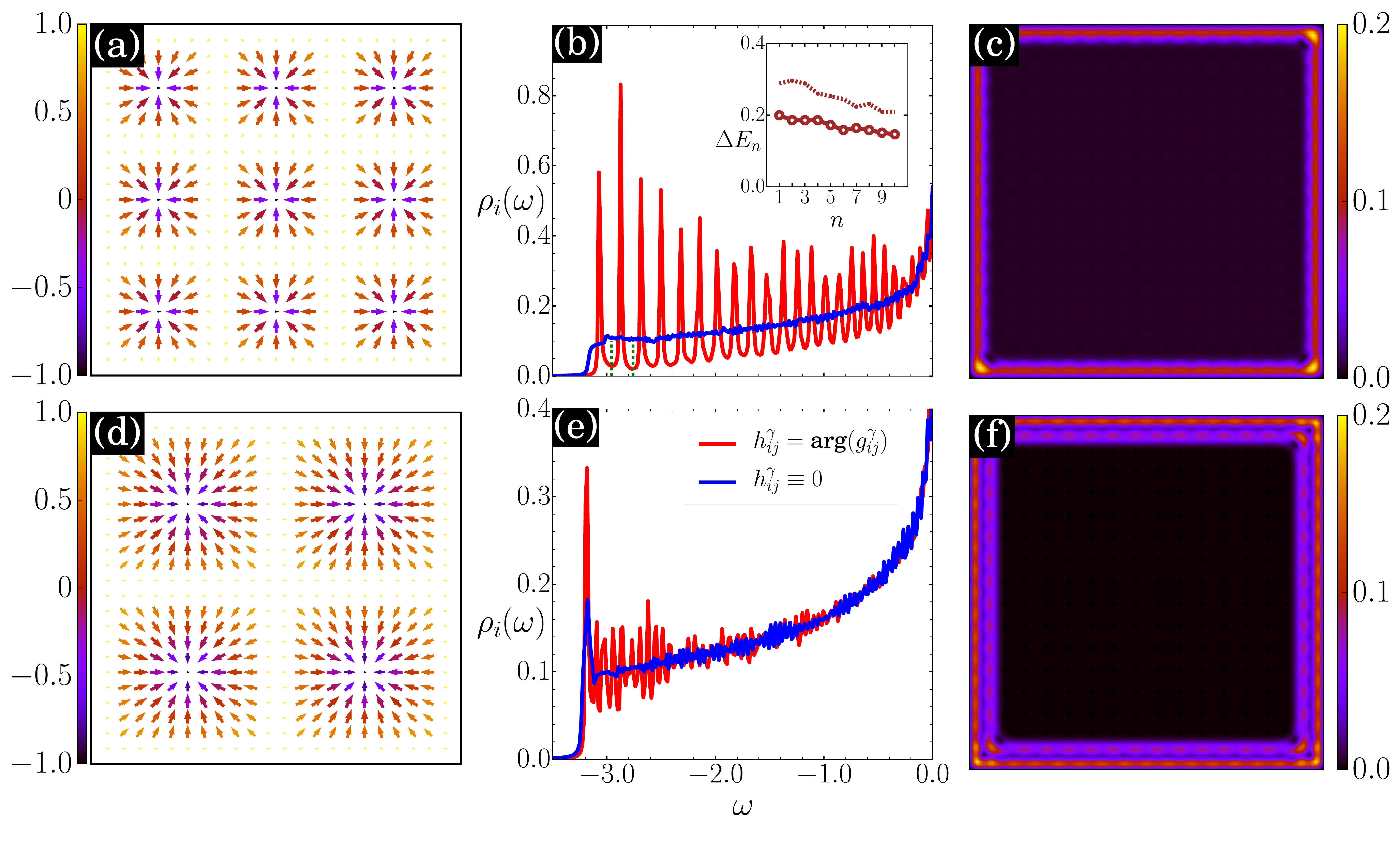}
\caption{{\bf Local gauge fields, Landau levels and boundary modes.} 
(a) Perfect skyrmion crystal configurations with, (a) $r_s = 1.5$ and (d) $r_s = 2.5$. (b) and (e) display LDOS in skyrmion cores, with (red lines) and without (blue lines) gauge fields, for configurations shown in (a) and (d), respectively. (c) and (f) display the LDOS maps for location of Fermi levels marked by green dotted lines, at $\omega = -2.95, -2.75$ in panel (b). One edge mode per filled Landau level is easily identified in panels (c) and (f). LDOS calculations are performed on lattice sizes $112 \times 112$ (for $r_s=1.5$) and $N=110 \times 110$ (for $r_s=2.5$).
We use $\alpha = 0.2$ for all calculations shown in this figure.
}
\label{fig6}
\end{figure*}

We underline the importance of the RZ mechanism for skyrmion formation in metals by presenting effects that are completely beyond the mechanisms that ignore electron itinerancy. We calculate local density of states, $\rho_i(\omega) = 1/N \sum_k |\psi^k_i|^2 \delta(\omega - E_k)$, where $\psi^k_i$ is the amplitude on site $i$ of the single particle eigenfunction corresponding to eigenvalue $E_k$ of the RDE Hamiltonian Eq. (\ref{Ham-RDE}). Lorentzian with broadening parameter $0.01$ is used to approximate the Dirac delta function. We find that the skyrmion textures in magnetization have strong implications for the electronic wavefunctions in this unusual metallic phase. 

We focus on the LDOS for sites located in skyrmion cores. In the sparse skyrmion case, there is a weak enhancement in LDOS near the band edge (see Fig. \ref{fig5}(c)). The effect becomes much pronounced for the packed skyrmion state. Furthermore, periodic modulations as a function of energy become clear (see Fig. \ref{fig5}(g)). Inset in Fig. \ref{fig5}(g) show the energy difference of two consecutive peaks, $\Delta E_n$, as a function of peak index. 
There are two possible interpretations of the spikes in LDOS. They can appear either due to the confinement effect, similar to those reported in metallic nanoislands and carbon nanotubes with defect \cite{Yang2005,Rastei2007}, or due to effective magnetic flux hidden in the gauge fields. We find a clear approach to disentangle these two effects. Ignoring the phases in the complex hopping parameters $g^{\gamma}_{ij}$ in the RDE Hamiltonian sets the gauge fields to zero and the resulting model with real hopping parameters contains pure confinement effects.
The results of LDOS calculation using $h^{\gamma}_{ij} \equiv 0$ in Eq. (\ref{Ham-RDE})(blue lines in Fig. \ref{fig5}(c), (g)) show that the periodic modulations vanish and only a single peak near the band edge survives. We plot lattice maps of LDOS for the energy fixed at peak location. The resulting maps 
display inhomogeneities, and a clear localization of electronic wavefunctions at skyrmion cores for the pSk state (see Fig. \ref{fig5}(h)).

The above analysis proves that, although the confinement effects are present due to change in the magnitude of $g^{\gamma}_{ij}$, the oscillations 
can only be explained by Landau level physics arising from effective magnetic flux hidden in complex $g^{\gamma}_{ij}$.
In order to confirm this, we set up a calculation where we reduce the disorder effects by designing ideal skyrmion lattice configurations. 
The elementary skyrmion unit is constructed by defining azimuthal and polar angles for localized spins as \cite{Tejo2018a}, $\phi_i = \pi + \tan^{-1}(y_i/x_i)$ and $\theta_i = 2 \tan^{-1}(r_s/r_i)e^{\beta(r_s-r_i)} ~ \Theta(2r_s-r_i)$, respectively. In the above, $x_i(y_i)$ denote the x (y) coordinate of the site $i$ located at distance $r_i$ from the skyrmion core site, $2r_s$ is the skyrmion radius and $\Theta$ denotes the Heaviside step function. We fix $\beta = 0.04$ to ensure similarity of ideal skyrmions with those obtained in HMC and effective Hamiltonian simulations. 

We show LDOS calculations for the ideal skyrmion crystals with $r_s = 1.5$ (Fig. \ref{fig6}(a)) and $r_s = 2.5$ (Fig. \ref{fig6}(d)). We obtain a very clear Landau level distribution for smaller skyrmions (Fig. \ref{fig6}(b)), whereas the Landau levels are not well separated for larger skyrmions. Therefore, smaller skyrmions generate stronger effective magnetic fields. 
Presence of disorder can further effect the separation of Landau levels, leading to an oscillatory behaviour only near the band edges as obtained for simulated skyrmion textures (see Fig. \ref{fig5}(g)). In Fig. \ref{fig6}(c) and (f) we show the LDOS maps calculated for the two locations of Fermi energies, corresponding to completely filled first and second Landau levels, marked by the vertical dotted lines in  Fig. \ref{fig6}(b). LDOS calculations explicitly show the presence of one (Fig. \ref{fig6}(c)) and two (Fig. \ref{fig6}(f)) edge modes in the two cases. Inset in Fig. \ref{fig6}(b) show the energy difference of two consecutive Landau levels as a function of Landau level index. Note that in continuum $\Delta E_n$ is independent of $n$, however in a tight binding model the energy dependence of the density of states leads to an $n$-dependence in $\Delta E_n$. The dashed line shows $\Delta E_n$ for a two dimensional tight-binding model with applied magnetic flux of strength $1/40$ flux quanta per square plaquette. This explicitly confirms that the gauge fields due to skyrmions play the same role as the external magnetic flux.
Since the features discussed in this section are unique to the RZ mechanism proposed in this work, they serve as testable predictions for the presence of the mechanism in thin films of magnetic metals. 

\section*{Conclusion}
Double exchange mechanism provides a basis for understanding ferromagnetism in a variety of metallic magnets. We have uncovered a new aspect associated with this classic text-book mechanism by including the effect of Rashba SOC in the DE model. We have presented an explicit demonstration of the existence of nanoscale skyrmions in a microscopic electronic model. The sparse and packed skyrmion states are shown to emerge from qualitatively distinct parent states.  
The circular patterns in the SSF are remarkably similar to those reported in the SANS experiments on Co-Zn-Mn alloys and MnSi \cite{Tonomura2012, Karube2016}. The corresponding real-space images, representative of fDW states, are also in agreement with the LTEM images on FeGe, Co-Zn-Mn and transition metal multilayers \cite{Pollard2017, Soumyanarayanan2017, Romming2013b,  Yu2011, Zhao2016, Tonomura2012,  Nagase2019}. Interestingly, domain walls junctions in the fDW state lead to local noncoplanar regions that emerge as nucleation centers for skyrmions in the presence of Zeeman field. For larger values of $\alpha$, the SQ spirals give way to pSk upon increasing $h_z$. The origin of these states lies in the anisotropy terms that become apparent in the effective Hamiltonian derived from the RDE model.
Existing DM based theories of skyrmion states promote the $(Q,Q)$ spiral at $h_z=0$ as the parent of skyrmion states. These theories were not able to explain the aforementioned experimental findings on a variety of metallic materials hosting skyrmions.
We have also shown, via Hall conductivity and Bott index calculations, that the skyrmion states induced by the RZ mechanism are examples of disordered topological metals. The local density of states bring out novel features of the skyrmion states in metals that are completely beyond the DM based pure spin models. 
We predict characteristic oscillations as a function of bias voltage in $dI/dV$ spectra, as experimental evidence for the RZ mechanism.
We believe that our discovery provides a conceptually consistent understanding of skyrmion formation in correlated magnetic metals. The possible tuning of skyrmion size down to nanoscale within the RZ mechanism will allow for an unprecedented data storage and processing capabilities.

\section*{Methods}
\subsection*{Classical Monte Carlo Simulations}
We simulate the spin Hamiltonian Eq. (\ref{eq:ESH}) via the conventional Classical Monte Carlo technique. In the zero field cooled protocol, the simulations begin in the paramagnetic phase with $h_z = 0$ and temperature is then lowered in discrete steps. To calculate the field dependence at low temperatures, which is the main focus of the study, the external field $h_z$ is increased in discrete steps. For a given value of $T$ and $h_z$, 
single spin updates are performed by proposing a new spin orientation, ${\bf S}'_i$, from a set of uniformly distributed points on the surface of a unit sphere. The new configuration is accepted based on the standard Metropolis algorithm. We use $\sim 5 \times 10^5$ Monte Carlo steps each for equilibration and averaging. For detailed exploration of parameter space we used lattice size $60 \times 60$, and the stability of results is ensured by simulating sizes up to $200 \times 200$  for some selected parameter values. In the field cooled protocol, the temperature is lowered in the presence of a finite external field.

\subsection*{Hybrid Monte Carlo Simulations}
The Hamiltonian Eq. (\ref{Ham-RDE}) belongs to a class of models with classical degrees of freedom coupled to electrons. Hybrid Monte Carlo simulations provide numerically exact approach for the study of such Hamiltonians. In this approach, the classical spin variables are updated according to Metropolis algorithm, however, the electronic Hamiltonian is diagonalized at each Monte Carlo step in order to compute the energy associated with a given spin configuration. The method is computationally expansive, and hence simulations are limited to lattices $\sim 100$ sites. For simulations on larger lattices, without compromising on the accuracy, we make use of the traveling cluster approximation (TCA) \cite{Kumar2006c, Mukherjee2015a}. In this method the exact diagonalization of the fermionic Hamiltonian is performed on a smaller cluster centered around the update site, and the cluster moves along with the update site.
The TCA simulations are performed on $24 \times 24$ lattice with periodic boundary conditions using an $8 \times 8$ cluster with open boundary conditions. The `CHEEVX' subroutine of the LAPACK library is used for diagonalization of the Hamiltonian. We use $\sim 10^3$ MC steps each for equilibration and averaging at each value of temperature and Zeeman field. Other details are same as in the classical Monte Carlo simulation method.
 
\subsection*{Hall Conductivity Calculation}
The Hall conductivity is computed by implementing the Kubo-Greenwood formula,
\begin{eqnarray}
\sigma_{xy}(E)= \frac{\textrm{i} e^2 \hbar}{N} \sum_{m} \sum_{n \neq m} (f(E_m) - f(E_n))\frac { \langle m |v_{x}| n \rangle \langle n |v_{y}|m \rangle} {(E_{m} - E_{n})^2 + \Gamma^2},
\label{eq:hall_cond} \nonumber
\end{eqnarray}
\noindent
where, $E_{m}$ is the eigenvalue corresponding to the eigenstate $|m\rangle$ and $f(E_{m})$ is the Fermi distribution function. The Lorentzian broadening parameter $\Gamma$ is taken to be $0.01$. The expression for the velocity operator along $\gamma$ direction, derived via the commutator with Hamiltonian of the position operator, is given by, 
$v_{\gamma} = \frac{\textrm{i}}{\hbar}[H_{\textrm{RDE}},r_{\gamma}] = -\frac{\textrm{i}}{\hbar} \sum_{\langle ij \rangle} [g^{\gamma}_{ij} d^\dagger_{i} d^{}_{j} - \textrm{H.c.}]$. The calculations are carried out at $T=0.01$.

\subsection*{Bott Index Calculation}
We compute the Bott index by following the standard algorithm described in literature \cite{Loring2010,Agarwala2017}. We find the details provided by Huang and Liu particularly useful for the stability of the numerical algorithm \cite{Huang2018, Huang2018a}. We use the idea of adding complementary projector and performing a singular value decomposition (SVD) as discussed by Huang and Liu. Diagonalization of complex non-symmetric matrices is performed using the `CGEES' subroutine, and the SVD using the `CGESVD' subroutine from the LAPACK package.

\section*{Acknowledgements}
We thank Goutam Sheet and Yogesh Singh for valuable discussions. We acknowledge the use of computing facility at IISER Mohali.


\begin{thebibliography}{54}%
\makeatletter
\providecommand \@ifxundefined [1]{%
 \@ifx{#1\undefined}
}%
\providecommand \@ifnum [1]{%
 \ifnum #1\expandafter \@firstoftwo
 \else \expandafter \@secondoftwo
 \fi
}%
\providecommand \@ifx [1]{%
 \ifx #1\expandafter \@firstoftwo
 \else \expandafter \@secondoftwo
 \fi
}%
\providecommand \natexlab [1]{#1}%
\providecommand \enquote  [1]{``#1''}%
\providecommand \bibnamefont  [1]{#1}%
\providecommand \bibfnamefont [1]{#1}%
\providecommand \citenamefont [1]{#1}%
\providecommand \href@noop [0]{\@secondoftwo}%
\providecommand \href [0]{\begingroup \@sanitize@url \@href}%
\providecommand \@href[1]{\@@startlink{#1}\@@href}%
\providecommand \@@href[1]{\endgroup#1\@@endlink}%
\providecommand \@sanitize@url [0]{\catcode `\\12\catcode `\$12\catcode
  `\&12\catcode `\#12\catcode `\^12\catcode `\_12\catcode `\%12\relax}%
\providecommand \@@startlink[1]{}%
\providecommand \@@endlink[0]{}%
\providecommand \url  [0]{\begingroup\@sanitize@url \@url }%
\providecommand \@url [1]{\endgroup\@href {#1}{\urlprefix }}%
\providecommand \urlprefix  [0]{URL }%
\providecommand \Eprint [0]{\href }%
\providecommand \doibase [0]{http://dx.doi.org/}%
\providecommand \selectlanguage [0]{\@gobble}%
\providecommand \bibinfo  [0]{\@secondoftwo}%
\providecommand \bibfield  [0]{\@secondoftwo}%
\providecommand \translation [1]{[#1]}%
\providecommand \BibitemOpen [0]{}%
\providecommand \bibitemStop [0]{}%
\providecommand \bibitemNoStop [0]{.\EOS\space}%
\providecommand \EOS [0]{\spacefactor3000\relax}%
\providecommand \BibitemShut  [1]{\csname bibitem#1\endcsname}%
\let\auto@bib@innerbib\@empty
\bibitem [{\citenamefont {Fert}\ \emph {et~al.}(2017)\citenamefont {Fert},
  \citenamefont {Reyren},\ and\ \citenamefont {Cros}}]{Fert2017}%
  \BibitemOpen
  \bibfield  {author} {\bibinfo {author} {\bibfnamefont {A.}~\bibnamefont
  {Fert}}, \bibinfo {author} {\bibfnamefont {N.}~\bibnamefont {Reyren}}, \ and\
  \bibinfo {author} {\bibfnamefont {V.}~\bibnamefont {Cros}},\ }\href {\doibase
  10.1038/natrevmats.2017.31} {\bibfield  {journal} {\bibinfo  {journal} {Nat.
  Rev. Mater.}\ }\textbf {\bibinfo {volume} {2}},\ \bibinfo {pages} {17031}
  (\bibinfo {year} {2017})}\BibitemShut {NoStop}%
\bibitem [{\citenamefont {Wiesendanger}(2016)}]{Wiesendanger2016}%
  \BibitemOpen
  \bibfield  {author} {\bibinfo {author} {\bibfnamefont {R.}~\bibnamefont
  {Wiesendanger}},\ }\href {\doibase 10.1038/natrevmats.2016.44} {\bibfield
  {journal} {\bibinfo  {journal} {Nat. Rev. Mater.}\ }\textbf {\bibinfo
  {volume} {1}},\ \bibinfo {pages} {16044} (\bibinfo {year}
  {2016})}\BibitemShut {NoStop}%
\bibitem [{\citenamefont {Fert}\ \emph {et~al.}(2013)\citenamefont {Fert},
  \citenamefont {Cros},\ and\ \citenamefont {Sampaio}}]{Fert2013}%
  \BibitemOpen
  \bibfield  {author} {\bibinfo {author} {\bibfnamefont {A.}~\bibnamefont
  {Fert}}, \bibinfo {author} {\bibfnamefont {V.}~\bibnamefont {Cros}}, \ and\
  \bibinfo {author} {\bibfnamefont {J.}~\bibnamefont {Sampaio}},\ }\href
  {\doibase 10.1038/nnano.2013.29} {\bibfield  {journal} {\bibinfo  {journal}
  {Nat. Nanotechnol.}\ }\textbf {\bibinfo {volume} {8}},\ \bibinfo {pages}
  {152} (\bibinfo {year} {2013})}\BibitemShut {NoStop}%
\bibitem [{\citenamefont {Nagaosa}\ and\ \citenamefont
  {Tokura}(2013)}]{Nagaosa2013b}%
  \BibitemOpen
  \bibfield  {author} {\bibinfo {author} {\bibfnamefont {N.}~\bibnamefont
  {Nagaosa}}\ and\ \bibinfo {author} {\bibfnamefont {Y.}~\bibnamefont
  {Tokura}},\ }\href {\doibase 10.1038/nnano.2013.243} {\bibfield  {journal}
  {\bibinfo  {journal} {Nat. Nanotechnol.}\ }\textbf {\bibinfo {volume} {8}},\
  \bibinfo {pages} {899} (\bibinfo {year} {2013})}\BibitemShut {NoStop}%
\bibitem [{\citenamefont {G{\"{o}}bel}\ \emph {et~al.}(2019)\citenamefont
  {G{\"{o}}bel}, \citenamefont {Mook}, \citenamefont {Henk},\ and\
  \citenamefont {Mertig}}]{Gobel2019}%
  \BibitemOpen
  \bibfield  {author} {\bibinfo {author} {\bibfnamefont {B.}~\bibnamefont
  {G{\"{o}}bel}}, \bibinfo {author} {\bibfnamefont {A.}~\bibnamefont {Mook}},
  \bibinfo {author} {\bibfnamefont {J.}~\bibnamefont {Henk}}, \ and\ \bibinfo
  {author} {\bibfnamefont {I.}~\bibnamefont {Mertig}},\ }\href {\doibase
  10.1103/PhysRevB.99.020405} {\bibfield  {journal} {\bibinfo  {journal} {Phys.
  Rev. B}\ }\textbf {\bibinfo {volume} {99}},\ \bibinfo {pages} {020405}
  (\bibinfo {year} {2019})}\BibitemShut {NoStop}%
\bibitem [{\citenamefont {Bogdanov}\ and\ \citenamefont
  {Panagopoulos}(2020)}]{Bogdanov2020}%
  \BibitemOpen
  \bibfield  {author} {\bibinfo {author} {\bibfnamefont {A.~N.}\ \bibnamefont
  {Bogdanov}}\ and\ \bibinfo {author} {\bibfnamefont {C.}~\bibnamefont
  {Panagopoulos}},\ }\href {\doibase 10.1063/PT.3.4431} {\bibfield  {journal}
  {\bibinfo  {journal} {Phys. Today}\ }\textbf {\bibinfo {volume} {73}},\
  \bibinfo {pages} {44} (\bibinfo {year} {2020})}\BibitemShut {NoStop}%
\bibitem [{\citenamefont {Dup{\'{e}}}\ \emph {et~al.}(2014)\citenamefont
  {Dup{\'{e}}}, \citenamefont {Hoffmann}, \citenamefont {Paillard},\ and\
  \citenamefont {Heinze}}]{Dupe2014}%
  \BibitemOpen
  \bibfield  {author} {\bibinfo {author} {\bibfnamefont {B.}~\bibnamefont
  {Dup{\'{e}}}}, \bibinfo {author} {\bibfnamefont {M.}~\bibnamefont
  {Hoffmann}}, \bibinfo {author} {\bibfnamefont {C.}~\bibnamefont {Paillard}},
  \ and\ \bibinfo {author} {\bibfnamefont {S.}~\bibnamefont {Heinze}},\ }\href
  {\doibase 10.1038/ncomms5030} {\bibfield  {journal} {\bibinfo  {journal}
  {Nat. Commun.}\ }\textbf {\bibinfo {volume} {5}},\ \bibinfo {pages} {4030}
  (\bibinfo {year} {2014})}\BibitemShut {NoStop}%
\bibitem [{\citenamefont {Pollard}\ \emph {et~al.}(2017)\citenamefont
  {Pollard}, \citenamefont {Garlow}, \citenamefont {Yu}, \citenamefont {Wang},
  \citenamefont {Zhu},\ and\ \citenamefont {Yang}}]{Pollard2017}%
  \BibitemOpen
  \bibfield  {author} {\bibinfo {author} {\bibfnamefont {S.~D.}\ \bibnamefont
  {Pollard}}, \bibinfo {author} {\bibfnamefont {J.~A.}\ \bibnamefont {Garlow}},
  \bibinfo {author} {\bibfnamefont {J.}~\bibnamefont {Yu}}, \bibinfo {author}
  {\bibfnamefont {Z.}~\bibnamefont {Wang}}, \bibinfo {author} {\bibfnamefont
  {Y.}~\bibnamefont {Zhu}}, \ and\ \bibinfo {author} {\bibfnamefont
  {H.}~\bibnamefont {Yang}},\ }\href {\doibase 10.1038/ncomms14761} {\bibfield
  {journal} {\bibinfo  {journal} {Nat. Commun.}\ }\textbf {\bibinfo {volume}
  {8}},\ \bibinfo {pages} {14761} (\bibinfo {year} {2017})}\BibitemShut
  {NoStop}%
\bibitem [{\citenamefont {Soumyanarayanan}\ \emph {et~al.}(2017)\citenamefont
  {Soumyanarayanan}, \citenamefont {Raju}, \citenamefont {{Gonzalez Oyarce}},
  \citenamefont {Tan}, \citenamefont {Im}, \citenamefont {Petrovi{\'{c}}},
  \citenamefont {Ho}, \citenamefont {Khoo}, \citenamefont {Tran}, \citenamefont
  {Gan}, \citenamefont {Ernult},\ and\ \citenamefont
  {Panagopoulos}}]{Soumyanarayanan2017}%
  \BibitemOpen
  \bibfield  {author} {\bibinfo {author} {\bibfnamefont {A.}~\bibnamefont
  {Soumyanarayanan}}, \bibinfo {author} {\bibfnamefont {M.}~\bibnamefont
  {Raju}}, \bibinfo {author} {\bibfnamefont {A.~L.}\ \bibnamefont {{Gonzalez
  Oyarce}}}, \bibinfo {author} {\bibfnamefont {A.~K.~C.}\ \bibnamefont {Tan}},
  \bibinfo {author} {\bibfnamefont {M.-Y.}\ \bibnamefont {Im}}, \bibinfo
  {author} {\bibfnamefont {A.}~\bibnamefont {Petrovi{\'{c}}}}, \bibinfo
  {author} {\bibfnamefont {P.}~\bibnamefont {Ho}}, \bibinfo {author}
  {\bibfnamefont {K.~H.}\ \bibnamefont {Khoo}}, \bibinfo {author}
  {\bibfnamefont {M.}~\bibnamefont {Tran}}, \bibinfo {author} {\bibfnamefont
  {C.~K.}\ \bibnamefont {Gan}}, \bibinfo {author} {\bibfnamefont
  {F.}~\bibnamefont {Ernult}}, \ and\ \bibinfo {author} {\bibfnamefont
  {C.}~\bibnamefont {Panagopoulos}},\ }\href {\doibase 10.1038/nmat4934}
  {\bibfield  {journal} {\bibinfo  {journal} {Nat. Mater.}\ }\textbf {\bibinfo
  {volume} {16}},\ \bibinfo {pages} {898} (\bibinfo {year} {2017})}\BibitemShut
  {NoStop}%
\bibitem [{\citenamefont {Romming}\ \emph {et~al.}(2013)\citenamefont
  {Romming}, \citenamefont {Hanneken}, \citenamefont {Menzel}, \citenamefont
  {Bickel}, \citenamefont {Wolter}, \citenamefont {von Bergmann}, \citenamefont
  {Kubetzka},\ and\ \citenamefont {Wiesendanger}}]{Romming2013b}%
  \BibitemOpen
  \bibfield  {author} {\bibinfo {author} {\bibfnamefont {N.}~\bibnamefont
  {Romming}}, \bibinfo {author} {\bibfnamefont {C.}~\bibnamefont {Hanneken}},
  \bibinfo {author} {\bibfnamefont {M.}~\bibnamefont {Menzel}}, \bibinfo
  {author} {\bibfnamefont {J.~E.}\ \bibnamefont {Bickel}}, \bibinfo {author}
  {\bibfnamefont {B.}~\bibnamefont {Wolter}}, \bibinfo {author} {\bibfnamefont
  {K.}~\bibnamefont {von Bergmann}}, \bibinfo {author} {\bibfnamefont
  {A.}~\bibnamefont {Kubetzka}}, \ and\ \bibinfo {author} {\bibfnamefont
  {R.}~\bibnamefont {Wiesendanger}},\ }\href {\doibase 10.1126/science.1240573}
  {\bibfield  {journal} {\bibinfo  {journal} {Science}\ }\textbf {\bibinfo
  {volume} {341}},\ \bibinfo {pages} {636} (\bibinfo {year}
  {2013})}\BibitemShut {NoStop}%
\bibitem [{\citenamefont {Yu}\ \emph {et~al.}(2012)\citenamefont {Yu},
  \citenamefont {Kanazawa}, \citenamefont {Zhang}, \citenamefont {Nagai},
  \citenamefont {Hara}, \citenamefont {Kimoto}, \citenamefont {Matsui},
  \citenamefont {Onose},\ and\ \citenamefont {Tokura}}]{Yu2012}%
  \BibitemOpen
  \bibfield  {author} {\bibinfo {author} {\bibfnamefont {X.~Z.}\ \bibnamefont
  {Yu}}, \bibinfo {author} {\bibfnamefont {N.}~\bibnamefont {Kanazawa}},
  \bibinfo {author} {\bibfnamefont {W.~Z.}\ \bibnamefont {Zhang}}, \bibinfo
  {author} {\bibfnamefont {T.}~\bibnamefont {Nagai}}, \bibinfo {author}
  {\bibfnamefont {T.}~\bibnamefont {Hara}}, \bibinfo {author} {\bibfnamefont
  {K.}~\bibnamefont {Kimoto}}, \bibinfo {author} {\bibfnamefont
  {Y.}~\bibnamefont {Matsui}}, \bibinfo {author} {\bibfnamefont
  {Y.}~\bibnamefont {Onose}}, \ and\ \bibinfo {author} {\bibfnamefont
  {Y.}~\bibnamefont {Tokura}},\ }\href {\doibase 10.1038/ncomms1990} {\bibfield
   {journal} {\bibinfo  {journal} {Nat. Commun.}\ }\textbf {\bibinfo {volume}
  {3}},\ \bibinfo {pages} {988} (\bibinfo {year} {2012})}\BibitemShut {NoStop}%
\bibitem [{\citenamefont {Yu}\ \emph {et~al.}(2011)\citenamefont {Yu},
  \citenamefont {Kanazawa}, \citenamefont {Onose}, \citenamefont {Kimoto},
  \citenamefont {Zhang}, \citenamefont {Ishiwata}, \citenamefont {Matsui},\
  and\ \citenamefont {Tokura}}]{Yu2011}%
  \BibitemOpen
  \bibfield  {author} {\bibinfo {author} {\bibfnamefont {X.~Z.}\ \bibnamefont
  {Yu}}, \bibinfo {author} {\bibfnamefont {N.}~\bibnamefont {Kanazawa}},
  \bibinfo {author} {\bibfnamefont {Y.}~\bibnamefont {Onose}}, \bibinfo
  {author} {\bibfnamefont {K.}~\bibnamefont {Kimoto}}, \bibinfo {author}
  {\bibfnamefont {W.~Z.}\ \bibnamefont {Zhang}}, \bibinfo {author}
  {\bibfnamefont {S.}~\bibnamefont {Ishiwata}}, \bibinfo {author}
  {\bibfnamefont {Y.}~\bibnamefont {Matsui}}, \ and\ \bibinfo {author}
  {\bibfnamefont {Y.}~\bibnamefont {Tokura}},\ }\href {\doibase
  10.1038/nmat2916} {\bibfield  {journal} {\bibinfo  {journal} {Nat. Mater.}\
  }\textbf {\bibinfo {volume} {10}},\ \bibinfo {pages} {106} (\bibinfo {year}
  {2011})}\BibitemShut {NoStop}%
\bibitem [{\citenamefont {Zhao}\ \emph {et~al.}(2016)\citenamefont {Zhao},
  \citenamefont {Jin}, \citenamefont {Wang}, \citenamefont {Du}, \citenamefont
  {Zang}, \citenamefont {Tian}, \citenamefont {Che},\ and\ \citenamefont
  {Zhang}}]{Zhao2016}%
  \BibitemOpen
  \bibfield  {author} {\bibinfo {author} {\bibfnamefont {X.}~\bibnamefont
  {Zhao}}, \bibinfo {author} {\bibfnamefont {C.}~\bibnamefont {Jin}}, \bibinfo
  {author} {\bibfnamefont {C.}~\bibnamefont {Wang}}, \bibinfo {author}
  {\bibfnamefont {H.}~\bibnamefont {Du}}, \bibinfo {author} {\bibfnamefont
  {J.}~\bibnamefont {Zang}}, \bibinfo {author} {\bibfnamefont {M.}~\bibnamefont
  {Tian}}, \bibinfo {author} {\bibfnamefont {R.}~\bibnamefont {Che}}, \ and\
  \bibinfo {author} {\bibfnamefont {Y.}~\bibnamefont {Zhang}},\ }\href
  {\doibase 10.1073/pnas.1600197113} {\bibfield  {journal} {\bibinfo  {journal}
  {Proc. Natl. Acad. Sci. U. S. A.}\ }\textbf {\bibinfo {volume} {113}},\
  \bibinfo {pages} {4918} (\bibinfo {year} {2016})}\BibitemShut {NoStop}%
\bibitem [{\citenamefont {Meyer}\ \emph {et~al.}(2019)\citenamefont {Meyer},
  \citenamefont {Perini}, \citenamefont {von Malottki}, \citenamefont
  {Kubetzka}, \citenamefont {Wiesendanger}, \citenamefont {von Bergmann},\ and\
  \citenamefont {Heinze}}]{Meyer2019}%
  \BibitemOpen
  \bibfield  {author} {\bibinfo {author} {\bibfnamefont {S.}~\bibnamefont
  {Meyer}}, \bibinfo {author} {\bibfnamefont {M.}~\bibnamefont {Perini}},
  \bibinfo {author} {\bibfnamefont {S.}~\bibnamefont {von Malottki}}, \bibinfo
  {author} {\bibfnamefont {A.}~\bibnamefont {Kubetzka}}, \bibinfo {author}
  {\bibfnamefont {R.}~\bibnamefont {Wiesendanger}}, \bibinfo {author}
  {\bibfnamefont {K.}~\bibnamefont {von Bergmann}}, \ and\ \bibinfo {author}
  {\bibfnamefont {S.}~\bibnamefont {Heinze}},\ }\href {\doibase
  10.1038/s41467-019-11831-4} {\bibfield  {journal} {\bibinfo  {journal} {Nat.
  Commun.}\ }\textbf {\bibinfo {volume} {10}},\ \bibinfo {pages} {3823}
  (\bibinfo {year} {2019})}\BibitemShut {NoStop}%
\bibitem [{\citenamefont {Tonomura}\ \emph {et~al.}(2012)\citenamefont
  {Tonomura}, \citenamefont {Yu}, \citenamefont {Yanagisawa}, \citenamefont
  {Matsuda}, \citenamefont {Onose}, \citenamefont {Kanazawa}, \citenamefont
  {Park},\ and\ \citenamefont {Tokura}}]{Tonomura2012}%
  \BibitemOpen
  \bibfield  {author} {\bibinfo {author} {\bibfnamefont {A.}~\bibnamefont
  {Tonomura}}, \bibinfo {author} {\bibfnamefont {X.}~\bibnamefont {Yu}},
  \bibinfo {author} {\bibfnamefont {K.}~\bibnamefont {Yanagisawa}}, \bibinfo
  {author} {\bibfnamefont {T.}~\bibnamefont {Matsuda}}, \bibinfo {author}
  {\bibfnamefont {Y.}~\bibnamefont {Onose}}, \bibinfo {author} {\bibfnamefont
  {N.}~\bibnamefont {Kanazawa}}, \bibinfo {author} {\bibfnamefont {H.~S.}\
  \bibnamefont {Park}}, \ and\ \bibinfo {author} {\bibfnamefont
  {Y.}~\bibnamefont {Tokura}},\ }\href {\doibase 10.1021/nl300073m} {\bibfield
  {journal} {\bibinfo  {journal} {Nano Lett.}\ }\textbf {\bibinfo {volume}
  {12}},\ \bibinfo {pages} {1673} (\bibinfo {year} {2012})}\BibitemShut
  {NoStop}%
\bibitem [{\citenamefont {Hirschberger}\ \emph {et~al.}(2019)\citenamefont
  {Hirschberger}, \citenamefont {Nakajima}, \citenamefont {Gao}, \citenamefont
  {Peng}, \citenamefont {Kikkawa}, \citenamefont {Kurumaji}, \citenamefont
  {Kriener}, \citenamefont {Yamasaki}, \citenamefont {Sagayama}, \citenamefont
  {Nakao}, \citenamefont {Ohishi}, \citenamefont {Kakurai}, \citenamefont
  {Taguchi}, \citenamefont {Yu}, \citenamefont {Arima},\ and\ \citenamefont
  {Tokura}}]{Hirschberger2019}%
  \BibitemOpen
  \bibfield  {author} {\bibinfo {author} {\bibfnamefont {M.}~\bibnamefont
  {Hirschberger}}, \bibinfo {author} {\bibfnamefont {T.}~\bibnamefont
  {Nakajima}}, \bibinfo {author} {\bibfnamefont {S.}~\bibnamefont {Gao}},
  \bibinfo {author} {\bibfnamefont {L.}~\bibnamefont {Peng}}, \bibinfo {author}
  {\bibfnamefont {A.}~\bibnamefont {Kikkawa}}, \bibinfo {author} {\bibfnamefont
  {T.}~\bibnamefont {Kurumaji}}, \bibinfo {author} {\bibfnamefont
  {M.}~\bibnamefont {Kriener}}, \bibinfo {author} {\bibfnamefont
  {Y.}~\bibnamefont {Yamasaki}}, \bibinfo {author} {\bibfnamefont
  {H.}~\bibnamefont {Sagayama}}, \bibinfo {author} {\bibfnamefont
  {H.}~\bibnamefont {Nakao}}, \bibinfo {author} {\bibfnamefont
  {K.}~\bibnamefont {Ohishi}}, \bibinfo {author} {\bibfnamefont
  {K.}~\bibnamefont {Kakurai}}, \bibinfo {author} {\bibfnamefont
  {Y.}~\bibnamefont {Taguchi}}, \bibinfo {author} {\bibfnamefont
  {X.}~\bibnamefont {Yu}}, \bibinfo {author} {\bibfnamefont {T.-h.}\
  \bibnamefont {Arima}}, \ and\ \bibinfo {author} {\bibfnamefont
  {Y.}~\bibnamefont {Tokura}},\ }\href {\doibase 10.1038/s41467-019-13675-4}
  {\bibfield  {journal} {\bibinfo  {journal} {Nat. Commun.}\ }\textbf {\bibinfo
  {volume} {10}},\ \bibinfo {pages} {5831} (\bibinfo {year}
  {2019})}\BibitemShut {NoStop}%
\bibitem [{\citenamefont {Jin}\ \emph {et~al.}(2017)\citenamefont {Jin},
  \citenamefont {Li}, \citenamefont {Kov{\'{a}}cs}, \citenamefont {Caron},
  \citenamefont {Zheng}, \citenamefont {Rybakov}, \citenamefont {Kiselev},
  \citenamefont {Du}, \citenamefont {Bl{\"{u}}gel}, \citenamefont {Tian},
  \citenamefont {Zhang}, \citenamefont {Farle},\ and\ \citenamefont
  {Dunin-Borkowski}}]{Jin2017}%
  \BibitemOpen
  \bibfield  {author} {\bibinfo {author} {\bibfnamefont {C.}~\bibnamefont
  {Jin}}, \bibinfo {author} {\bibfnamefont {Z.-A.}\ \bibnamefont {Li}},
  \bibinfo {author} {\bibfnamefont {A.}~\bibnamefont {Kov{\'{a}}cs}}, \bibinfo
  {author} {\bibfnamefont {J.}~\bibnamefont {Caron}}, \bibinfo {author}
  {\bibfnamefont {F.}~\bibnamefont {Zheng}}, \bibinfo {author} {\bibfnamefont
  {F.~N.}\ \bibnamefont {Rybakov}}, \bibinfo {author} {\bibfnamefont {N.~S.}\
  \bibnamefont {Kiselev}}, \bibinfo {author} {\bibfnamefont {H.}~\bibnamefont
  {Du}}, \bibinfo {author} {\bibfnamefont {S.}~\bibnamefont {Bl{\"{u}}gel}},
  \bibinfo {author} {\bibfnamefont {M.}~\bibnamefont {Tian}}, \bibinfo {author}
  {\bibfnamefont {Y.}~\bibnamefont {Zhang}}, \bibinfo {author} {\bibfnamefont
  {M.}~\bibnamefont {Farle}}, \ and\ \bibinfo {author} {\bibfnamefont {R.~E.}\
  \bibnamefont {Dunin-Borkowski}},\ }\href {\doibase 10.1038/ncomms15569}
  {\bibfield  {journal} {\bibinfo  {journal} {Nat. Commun.}\ }\textbf {\bibinfo
  {volume} {8}},\ \bibinfo {pages} {15569} (\bibinfo {year}
  {2017})}\BibitemShut {NoStop}%
\bibitem [{\citenamefont {Karube}\ \emph {et~al.}(2016)\citenamefont {Karube},
  \citenamefont {White}, \citenamefont {Reynolds}, \citenamefont {Gavilano},
  \citenamefont {Oike}, \citenamefont {Kikkawa}, \citenamefont {Kagawa},
  \citenamefont {Tokunaga}, \citenamefont {R{\o}nnow}, \citenamefont {Tokura},\
  and\ \citenamefont {Taguchi}}]{Karube2016}%
  \BibitemOpen
  \bibfield  {author} {\bibinfo {author} {\bibfnamefont {K.}~\bibnamefont
  {Karube}}, \bibinfo {author} {\bibfnamefont {J.~S.}\ \bibnamefont {White}},
  \bibinfo {author} {\bibfnamefont {N.}~\bibnamefont {Reynolds}}, \bibinfo
  {author} {\bibfnamefont {J.~L.}\ \bibnamefont {Gavilano}}, \bibinfo {author}
  {\bibfnamefont {H.}~\bibnamefont {Oike}}, \bibinfo {author} {\bibfnamefont
  {A.}~\bibnamefont {Kikkawa}}, \bibinfo {author} {\bibfnamefont
  {F.}~\bibnamefont {Kagawa}}, \bibinfo {author} {\bibfnamefont
  {Y.}~\bibnamefont {Tokunaga}}, \bibinfo {author} {\bibfnamefont {H.~M.}\
  \bibnamefont {R{\o}nnow}}, \bibinfo {author} {\bibfnamefont {Y.}~\bibnamefont
  {Tokura}}, \ and\ \bibinfo {author} {\bibfnamefont {Y.}~\bibnamefont
  {Taguchi}},\ }\href {\doibase 10.1038/nmat4752} {\bibfield  {journal}
  {\bibinfo  {journal} {Nat. Mater.}\ }\textbf {\bibinfo {volume} {15}},\
  \bibinfo {pages} {1237} (\bibinfo {year} {2016})}\BibitemShut {NoStop}%
\bibitem [{\citenamefont {M{\"{u}}hlbauer}\ \emph {et~al.}(2009)\citenamefont
  {M{\"{u}}hlbauer}, \citenamefont {Binz}, \citenamefont {Jonietz},
  \citenamefont {Pfleiderer}, \citenamefont {Rosch}, \citenamefont {Neubauer},
  \citenamefont {Georgii},\ and\ \citenamefont {B{\"{o}}ni}}]{Muhlbauer2009b}%
  \BibitemOpen
  \bibfield  {author} {\bibinfo {author} {\bibfnamefont {S.}~\bibnamefont
  {M{\"{u}}hlbauer}}, \bibinfo {author} {\bibfnamefont {B.}~\bibnamefont
  {Binz}}, \bibinfo {author} {\bibfnamefont {F.}~\bibnamefont {Jonietz}},
  \bibinfo {author} {\bibfnamefont {C.}~\bibnamefont {Pfleiderer}}, \bibinfo
  {author} {\bibfnamefont {A.}~\bibnamefont {Rosch}}, \bibinfo {author}
  {\bibfnamefont {A.}~\bibnamefont {Neubauer}}, \bibinfo {author}
  {\bibfnamefont {R.}~\bibnamefont {Georgii}}, \ and\ \bibinfo {author}
  {\bibfnamefont {P.}~\bibnamefont {B{\"{o}}ni}},\ }\href {\doibase
  10.1126/science.1166767} {\bibfield  {journal} {\bibinfo  {journal}
  {Science}\ }\textbf {\bibinfo {volume} {323}},\ \bibinfo {pages} {915}
  (\bibinfo {year} {2009})}\BibitemShut {NoStop}%
\bibitem [{\citenamefont {Song}\ \emph {et~al.}(2020)\citenamefont {Song},
  \citenamefont {Jeong}, \citenamefont {Pan}, \citenamefont {Zhang},
  \citenamefont {Xia}, \citenamefont {Cha}, \citenamefont {Park}, \citenamefont
  {Kim}, \citenamefont {Finizio}, \citenamefont {Raabe}, \citenamefont {Chang},
  \citenamefont {Zhou}, \citenamefont {Zhao}, \citenamefont {Kang},
  \citenamefont {Ju},\ and\ \citenamefont {Woo}}]{Song2020}%
  \BibitemOpen
  \bibfield  {author} {\bibinfo {author} {\bibfnamefont {K.~M.}\ \bibnamefont
  {Song}}, \bibinfo {author} {\bibfnamefont {J.-S.}\ \bibnamefont {Jeong}},
  \bibinfo {author} {\bibfnamefont {B.}~\bibnamefont {Pan}}, \bibinfo {author}
  {\bibfnamefont {X.}~\bibnamefont {Zhang}}, \bibinfo {author} {\bibfnamefont
  {J.}~\bibnamefont {Xia}}, \bibinfo {author} {\bibfnamefont {S.}~\bibnamefont
  {Cha}}, \bibinfo {author} {\bibfnamefont {T.-E.}\ \bibnamefont {Park}},
  \bibinfo {author} {\bibfnamefont {K.}~\bibnamefont {Kim}}, \bibinfo {author}
  {\bibfnamefont {S.}~\bibnamefont {Finizio}}, \bibinfo {author} {\bibfnamefont
  {J.}~\bibnamefont {Raabe}}, \bibinfo {author} {\bibfnamefont
  {J.}~\bibnamefont {Chang}}, \bibinfo {author} {\bibfnamefont
  {Y.}~\bibnamefont {Zhou}}, \bibinfo {author} {\bibfnamefont {W.}~\bibnamefont
  {Zhao}}, \bibinfo {author} {\bibfnamefont {W.}~\bibnamefont {Kang}}, \bibinfo
  {author} {\bibfnamefont {H.}~\bibnamefont {Ju}}, \ and\ \bibinfo {author}
  {\bibfnamefont {S.}~\bibnamefont {Woo}},\ }\href {\doibase
  10.1038/s41928-020-0385-0} {\bibfield  {journal} {\bibinfo  {journal} {Nat.
  Electron.}\ }\textbf {\bibinfo {volume} {3}},\ \bibinfo {pages} {148}
  (\bibinfo {year} {2020})}\BibitemShut {NoStop}%
\bibitem [{\citenamefont {Sampaio}\ \emph {et~al.}(2013)\citenamefont
  {Sampaio}, \citenamefont {Cros}, \citenamefont {Rohart}, \citenamefont
  {Thiaville},\ and\ \citenamefont {Fert}}]{Sampaio2013}%
  \BibitemOpen
  \bibfield  {author} {\bibinfo {author} {\bibfnamefont {J.}~\bibnamefont
  {Sampaio}}, \bibinfo {author} {\bibfnamefont {V.}~\bibnamefont {Cros}},
  \bibinfo {author} {\bibfnamefont {S.}~\bibnamefont {Rohart}}, \bibinfo
  {author} {\bibfnamefont {A.}~\bibnamefont {Thiaville}}, \ and\ \bibinfo
  {author} {\bibfnamefont {A.}~\bibnamefont {Fert}},\ }\href {\doibase
  10.1038/nnano.2013.210} {\bibfield  {journal} {\bibinfo  {journal} {Nat.
  Nanotechnol.}\ }\textbf {\bibinfo {volume} {8}},\ \bibinfo {pages} {839}
  (\bibinfo {year} {2013})}\BibitemShut {NoStop}%
\bibitem [{\citenamefont {Nayak}\ \emph {et~al.}(2017)\citenamefont {Nayak},
  \citenamefont {Kumar}, \citenamefont {Ma}, \citenamefont {Werner},
  \citenamefont {Pippel}, \citenamefont {Sahoo}, \citenamefont {Damay},
  \citenamefont {R{\"{o}}{\ss}ler}, \citenamefont {Felser},\ and\ \citenamefont
  {Parkin}}]{Nayak2017}%
  \BibitemOpen
  \bibfield  {author} {\bibinfo {author} {\bibfnamefont {A.~K.}\ \bibnamefont
  {Nayak}}, \bibinfo {author} {\bibfnamefont {V.}~\bibnamefont {Kumar}},
  \bibinfo {author} {\bibfnamefont {T.}~\bibnamefont {Ma}}, \bibinfo {author}
  {\bibfnamefont {P.}~\bibnamefont {Werner}}, \bibinfo {author} {\bibfnamefont
  {E.}~\bibnamefont {Pippel}}, \bibinfo {author} {\bibfnamefont
  {R.}~\bibnamefont {Sahoo}}, \bibinfo {author} {\bibfnamefont
  {F.}~\bibnamefont {Damay}}, \bibinfo {author} {\bibfnamefont {U.~K.}\
  \bibnamefont {R{\"{o}}{\ss}ler}}, \bibinfo {author} {\bibfnamefont
  {C.}~\bibnamefont {Felser}}, \ and\ \bibinfo {author} {\bibfnamefont
  {S.~S.~P.}\ \bibnamefont {Parkin}},\ }\href {\doibase 10.1038/nature23466}
  {\bibfield  {journal} {\bibinfo  {journal} {Nature}\ }\textbf {\bibinfo
  {volume} {548}},\ \bibinfo {pages} {561} (\bibinfo {year}
  {2017})}\BibitemShut {NoStop}%
\bibitem [{\citenamefont {Pfleiderer}\ \emph {et~al.}(2004)\citenamefont
  {Pfleiderer}, \citenamefont {Reznik}, \citenamefont {Pintschovius},
  \citenamefont {L{\"{o}}hneysen}, \citenamefont {Garst},\ and\ \citenamefont
  {Rosch}}]{Pfleiderer2004}%
  \BibitemOpen
  \bibfield  {author} {\bibinfo {author} {\bibfnamefont {C.}~\bibnamefont
  {Pfleiderer}}, \bibinfo {author} {\bibfnamefont {D.}~\bibnamefont {Reznik}},
  \bibinfo {author} {\bibfnamefont {L.}~\bibnamefont {Pintschovius}}, \bibinfo
  {author} {\bibfnamefont {H.~v.}\ \bibnamefont {L{\"{o}}hneysen}}, \bibinfo
  {author} {\bibfnamefont {M.}~\bibnamefont {Garst}}, \ and\ \bibinfo {author}
  {\bibfnamefont {A.}~\bibnamefont {Rosch}},\ }\href {\doibase
  10.1038/nature02232} {\bibfield  {journal} {\bibinfo  {journal} {Nature}\
  }\textbf {\bibinfo {volume} {427}},\ \bibinfo {pages} {227} (\bibinfo {year}
  {2004})}\BibitemShut {NoStop}%
\bibitem [{\citenamefont {Jena}\ \emph {et~al.}(2020)\citenamefont {Jena},
  \citenamefont {G{\"{o}}bel}, \citenamefont {Ma}, \citenamefont {Kumar},
  \citenamefont {Saha}, \citenamefont {Mertig}, \citenamefont {Felser},\ and\
  \citenamefont {Parkin}}]{Jena2020}%
  \BibitemOpen
  \bibfield  {author} {\bibinfo {author} {\bibfnamefont {J.}~\bibnamefont
  {Jena}}, \bibinfo {author} {\bibfnamefont {B.}~\bibnamefont {G{\"{o}}bel}},
  \bibinfo {author} {\bibfnamefont {T.}~\bibnamefont {Ma}}, \bibinfo {author}
  {\bibfnamefont {V.}~\bibnamefont {Kumar}}, \bibinfo {author} {\bibfnamefont
  {R.}~\bibnamefont {Saha}}, \bibinfo {author} {\bibfnamefont {I.}~\bibnamefont
  {Mertig}}, \bibinfo {author} {\bibfnamefont {C.}~\bibnamefont {Felser}}, \
  and\ \bibinfo {author} {\bibfnamefont {S.~S.~P.}\ \bibnamefont {Parkin}},\
  }\href {\doibase 10.1038/s41467-020-14925-6} {\bibfield  {journal} {\bibinfo
  {journal} {Nat. Commun.}\ }\textbf {\bibinfo {volume} {11}},\ \bibinfo
  {pages} {1115} (\bibinfo {year} {2020})}\BibitemShut {NoStop}%
\bibitem [{\citenamefont {Hsu}\ \emph {et~al.}(2018)\citenamefont {Hsu},
  \citenamefont {R{\'{o}}zsa}, \citenamefont {Finco}, \citenamefont {Schmidt},
  \citenamefont {Palot{\'{a}}s}, \citenamefont {Vedmedenko}, \citenamefont
  {Udvardi}, \citenamefont {Szunyogh}, \citenamefont {Kubetzka}, \citenamefont
  {von Bergmann},\ and\ \citenamefont {Wiesendanger}}]{Hsu2018}%
  \BibitemOpen
  \bibfield  {author} {\bibinfo {author} {\bibfnamefont {P.-J.}\ \bibnamefont
  {Hsu}}, \bibinfo {author} {\bibfnamefont {L.}~\bibnamefont {R{\'{o}}zsa}},
  \bibinfo {author} {\bibfnamefont {A.}~\bibnamefont {Finco}}, \bibinfo
  {author} {\bibfnamefont {L.}~\bibnamefont {Schmidt}}, \bibinfo {author}
  {\bibfnamefont {K.}~\bibnamefont {Palot{\'{a}}s}}, \bibinfo {author}
  {\bibfnamefont {E.}~\bibnamefont {Vedmedenko}}, \bibinfo {author}
  {\bibfnamefont {L.}~\bibnamefont {Udvardi}}, \bibinfo {author} {\bibfnamefont
  {L.}~\bibnamefont {Szunyogh}}, \bibinfo {author} {\bibfnamefont
  {A.}~\bibnamefont {Kubetzka}}, \bibinfo {author} {\bibfnamefont
  {K.}~\bibnamefont {von Bergmann}}, \ and\ \bibinfo {author} {\bibfnamefont
  {R.}~\bibnamefont {Wiesendanger}},\ }\href {\doibase
  10.1038/s41467-018-04015-z} {\bibfield  {journal} {\bibinfo  {journal} {Nat.
  Commun.}\ }\textbf {\bibinfo {volume} {9}},\ \bibinfo {pages} {1571}
  (\bibinfo {year} {2018})}\BibitemShut {NoStop}%
\bibitem [{\citenamefont {Yu}\ \emph {et~al.}(2018)\citenamefont {Yu},
  \citenamefont {Koshibae}, \citenamefont {Tokunaga}, \citenamefont {Shibata},
  \citenamefont {Taguchi}, \citenamefont {Nagaosa},\ and\ \citenamefont
  {Tokura}}]{Yu2018}%
  \BibitemOpen
  \bibfield  {author} {\bibinfo {author} {\bibfnamefont {X.~Z.}\ \bibnamefont
  {Yu}}, \bibinfo {author} {\bibfnamefont {W.}~\bibnamefont {Koshibae}},
  \bibinfo {author} {\bibfnamefont {Y.}~\bibnamefont {Tokunaga}}, \bibinfo
  {author} {\bibfnamefont {K.}~\bibnamefont {Shibata}}, \bibinfo {author}
  {\bibfnamefont {Y.}~\bibnamefont {Taguchi}}, \bibinfo {author} {\bibfnamefont
  {N.}~\bibnamefont {Nagaosa}}, \ and\ \bibinfo {author} {\bibfnamefont
  {Y.}~\bibnamefont {Tokura}},\ }\href {\doibase 10.1038/s41586-018-0745-3}
  {\bibfield  {journal} {\bibinfo  {journal} {Nature}\ }\textbf {\bibinfo
  {volume} {564}},\ \bibinfo {pages} {95} (\bibinfo {year} {2018})}\BibitemShut
  {NoStop}%
\bibitem [{\citenamefont {Nagase}\ \emph {et~al.}(2019)\citenamefont {Nagase},
  \citenamefont {Komatsu}, \citenamefont {So}, \citenamefont {Ishida},
  \citenamefont {Yoshida}, \citenamefont {Kawaguchi}, \citenamefont {Tanaka},
  \citenamefont {Saitoh}, \citenamefont {Ikarashi}, \citenamefont {Kuwahara},\
  and\ \citenamefont {Nagao}}]{Nagase2019}%
  \BibitemOpen
  \bibfield  {author} {\bibinfo {author} {\bibfnamefont {T.}~\bibnamefont
  {Nagase}}, \bibinfo {author} {\bibfnamefont {M.}~\bibnamefont {Komatsu}},
  \bibinfo {author} {\bibfnamefont {Y.~G.}\ \bibnamefont {So}}, \bibinfo
  {author} {\bibfnamefont {T.}~\bibnamefont {Ishida}}, \bibinfo {author}
  {\bibfnamefont {H.}~\bibnamefont {Yoshida}}, \bibinfo {author} {\bibfnamefont
  {Y.}~\bibnamefont {Kawaguchi}}, \bibinfo {author} {\bibfnamefont
  {Y.}~\bibnamefont {Tanaka}}, \bibinfo {author} {\bibfnamefont
  {K.}~\bibnamefont {Saitoh}}, \bibinfo {author} {\bibfnamefont
  {N.}~\bibnamefont {Ikarashi}}, \bibinfo {author} {\bibfnamefont
  {M.}~\bibnamefont {Kuwahara}}, \ and\ \bibinfo {author} {\bibfnamefont
  {M.}~\bibnamefont {Nagao}},\ }\href {\doibase 10.1103/PhysRevLett.123.137203}
  {\bibfield  {journal} {\bibinfo  {journal} {Phys. Rev. Lett.}\ }\textbf
  {\bibinfo {volume} {123}},\ \bibinfo {pages} {137203} (\bibinfo {year}
  {2019})}\BibitemShut {NoStop}%
\bibitem [{\citenamefont {R{\"{o}}{\ss}ler}\ \emph {et~al.}(2006)\citenamefont
  {R{\"{o}}{\ss}ler}, \citenamefont {Bogdanov},\ and\ \citenamefont
  {Pfleiderer}}]{Rossler2006}%
  \BibitemOpen
  \bibfield  {author} {\bibinfo {author} {\bibfnamefont {U.~K.}\ \bibnamefont
  {R{\"{o}}{\ss}ler}}, \bibinfo {author} {\bibfnamefont {A.~N.}\ \bibnamefont
  {Bogdanov}}, \ and\ \bibinfo {author} {\bibfnamefont {C.}~\bibnamefont
  {Pfleiderer}},\ }\href {\doibase 10.1038/nature05056} {\bibfield  {journal}
  {\bibinfo  {journal} {Nature}\ }\textbf {\bibinfo {volume} {442}},\ \bibinfo
  {pages} {797} (\bibinfo {year} {2006})}\BibitemShut {NoStop}%
\bibitem [{\citenamefont {Chen}\ \emph {et~al.}(2016)\citenamefont {Chen},
  \citenamefont {Zhang},\ and\ \citenamefont {Liu}}]{Chen2016}%
  \BibitemOpen
  \bibfield  {author} {\bibinfo {author} {\bibfnamefont {J.~P.}\ \bibnamefont
  {Chen}}, \bibinfo {author} {\bibfnamefont {D.-W.}\ \bibnamefont {Zhang}}, \
  and\ \bibinfo {author} {\bibfnamefont {J.~M.}\ \bibnamefont {Liu}},\ }\href
  {\doibase 10.1038/srep29126} {\bibfield  {journal} {\bibinfo  {journal} {Sci.
  Rep.}\ }\textbf {\bibinfo {volume} {6}},\ \bibinfo {pages} {29126} (\bibinfo
  {year} {2016})}\BibitemShut {NoStop}%
\bibitem [{\citenamefont {Mohanta}\ \emph {et~al.}(2019)\citenamefont
  {Mohanta}, \citenamefont {Dagotto},\ and\ \citenamefont
  {Okamoto}}]{Mohanta2019}%
  \BibitemOpen
  \bibfield  {author} {\bibinfo {author} {\bibfnamefont {N.}~\bibnamefont
  {Mohanta}}, \bibinfo {author} {\bibfnamefont {E.}~\bibnamefont {Dagotto}}, \
  and\ \bibinfo {author} {\bibfnamefont {S.}~\bibnamefont {Okamoto}},\ }\href
  {\doibase 10.1103/PhysRevB.100.064429} {\bibfield  {journal} {\bibinfo
  {journal} {Phys. Rev. B}\ }\textbf {\bibinfo {volume} {100}} (\bibinfo {year}
  {2019}),\ 10.1103/PhysRevB.100.064429}\BibitemShut {NoStop}%
\bibitem [{\citenamefont {Zang}\ \emph {et~al.}(2011)\citenamefont {Zang},
  \citenamefont {Mostovoy}, \citenamefont {Han},\ and\ \citenamefont
  {Nagaosa}}]{Zang2011a}%
  \BibitemOpen
  \bibfield  {author} {\bibinfo {author} {\bibfnamefont {J.}~\bibnamefont
  {Zang}}, \bibinfo {author} {\bibfnamefont {M.}~\bibnamefont {Mostovoy}},
  \bibinfo {author} {\bibfnamefont {J.~H.}\ \bibnamefont {Han}}, \ and\
  \bibinfo {author} {\bibfnamefont {N.}~\bibnamefont {Nagaosa}},\ }\href
  {\doibase 10.1103/PhysRevLett.107.136804} {\bibfield  {journal} {\bibinfo
  {journal} {Phys. Rev. Lett.}\ }\textbf {\bibinfo {volume} {107}},\ \bibinfo
  {pages} {136804} (\bibinfo {year} {2011})}\BibitemShut {NoStop}%
\bibitem [{\citenamefont {Iwasaki}\ \emph {et~al.}(2014)\citenamefont
  {Iwasaki}, \citenamefont {Beekman},\ and\ \citenamefont
  {Nagaosa}}]{Iwasaki2014}%
  \BibitemOpen
  \bibfield  {author} {\bibinfo {author} {\bibfnamefont {J.}~\bibnamefont
  {Iwasaki}}, \bibinfo {author} {\bibfnamefont {A.~J.}\ \bibnamefont
  {Beekman}}, \ and\ \bibinfo {author} {\bibfnamefont {N.}~\bibnamefont
  {Nagaosa}},\ }\href {\doibase 10.1103/PhysRevB.89.064412} {\bibfield
  {journal} {\bibinfo  {journal} {Phys. Rev. B}\ }\textbf {\bibinfo {volume}
  {89}},\ \bibinfo {pages} {064412} (\bibinfo {year} {2014})}\BibitemShut
  {NoStop}%
\bibitem [{\citenamefont {Ozawa}\ \emph {et~al.}(2017)\citenamefont {Ozawa},
  \citenamefont {Hayami},\ and\ \citenamefont {Motome}}]{Ozawa2017a}%
  \BibitemOpen
  \bibfield  {author} {\bibinfo {author} {\bibfnamefont {R.}~\bibnamefont
  {Ozawa}}, \bibinfo {author} {\bibfnamefont {S.}~\bibnamefont {Hayami}}, \
  and\ \bibinfo {author} {\bibfnamefont {Y.}~\bibnamefont {Motome}},\ }\href
  {\doibase 10.1103/PhysRevLett.118.147205} {\bibfield  {journal} {\bibinfo
  {journal} {Phys. Rev. Lett.}\ }\textbf {\bibinfo {volume} {118}},\ \bibinfo
  {pages} {147205} (\bibinfo {year} {2017})}\BibitemShut {NoStop}%
\bibitem [{\citenamefont {Wang}\ \emph {et~al.}(2020)\citenamefont {Wang},
  \citenamefont {Su}, \citenamefont {Lin},\ and\ \citenamefont
  {Batista}}]{Wang2020}%
  \BibitemOpen
  \bibfield  {author} {\bibinfo {author} {\bibfnamefont {Z.}~\bibnamefont
  {Wang}}, \bibinfo {author} {\bibfnamefont {Y.}~\bibnamefont {Su}}, \bibinfo
  {author} {\bibfnamefont {S.-Z.}\ \bibnamefont {Lin}}, \ and\ \bibinfo
  {author} {\bibfnamefont {C.~D.}\ \bibnamefont {Batista}},\ }\href {\doibase
  10.1103/PhysRevLett.124.207201} {\bibfield  {journal} {\bibinfo  {journal}
  {Phys. Rev. Lett.}\ }\textbf {\bibinfo {volume} {124}},\ \bibinfo {pages}
  {207201} (\bibinfo {year} {2020})}\BibitemShut {NoStop}%
\bibitem [{\citenamefont {Zener}(1951)}]{Zener1951a}%
  \BibitemOpen
  \bibfield  {author} {\bibinfo {author} {\bibfnamefont {C.}~\bibnamefont
  {Zener}},\ }\href {\doibase 10.1103/PhysRev.82.403} {\bibfield  {journal}
  {\bibinfo  {journal} {Phys. Rev.}\ }\textbf {\bibinfo {volume} {82}},\
  \bibinfo {pages} {403} (\bibinfo {year} {1951})}\BibitemShut {NoStop}%
\bibitem [{\citenamefont {Anderson}\ and\ \citenamefont
  {Hasegawa}(1955)}]{Anderson1955}%
  \BibitemOpen
  \bibfield  {author} {\bibinfo {author} {\bibfnamefont {P.~W.}\ \bibnamefont
  {Anderson}}\ and\ \bibinfo {author} {\bibfnamefont {H.}~\bibnamefont
  {Hasegawa}},\ }\href {\doibase 10.1103/PhysRev.100.675} {\bibfield  {journal}
  {\bibinfo  {journal} {Phys. Rev.}\ }\textbf {\bibinfo {volume} {100}},\
  \bibinfo {pages} {675} (\bibinfo {year} {1955})}\BibitemShut {NoStop}%
\bibitem [{\citenamefont {de~Gennes}(1960)}]{deGennes1960a}%
  \BibitemOpen
  \bibfield  {author} {\bibinfo {author} {\bibfnamefont {P.~G.}\ \bibnamefont
  {de~Gennes}},\ }\href {\doibase 10.1103/PhysRev.118.141} {\bibfield
  {journal} {\bibinfo  {journal} {Phys. Rev.}\ }\textbf {\bibinfo {volume}
  {118}},\ \bibinfo {pages} {141} (\bibinfo {year} {1960})}\BibitemShut
  {NoStop}%
\bibitem [{\citenamefont {Dagotto}(2002)}]{Dagotto2002}%
  \BibitemOpen
  \bibfield  {author} {\bibinfo {author} {\bibfnamefont {E.}~\bibnamefont
  {Dagotto}},\ }\href@noop {} {\emph {\bibinfo {title} {{Nanoscale Phase
  Separation and Collosal Magnetoresistance}}}}\ (\bibinfo  {publisher}
  {Springer Berlin/Heidelberg},\ \bibinfo {address} {Berlin},\ \bibinfo {year}
  {2002})\BibitemShut {NoStop}%
\bibitem [{\citenamefont {Pradhan}\ and\ \citenamefont
  {Das}(2017)}]{Pradhan2017}%
  \BibitemOpen
  \bibfield  {author} {\bibinfo {author} {\bibfnamefont {K.}~\bibnamefont
  {Pradhan}}\ and\ \bibinfo {author} {\bibfnamefont {S.~K.}\ \bibnamefont
  {Das}},\ }\href {\doibase 10.1038/s41598-017-09729-6} {\bibfield  {journal}
  {\bibinfo  {journal} {Sci. Rep.}\ }\textbf {\bibinfo {volume} {7}},\ \bibinfo
  {pages} {9603} (\bibinfo {year} {2017})}\BibitemShut {NoStop}%
\bibitem [{\citenamefont {Yaouanc}\ \emph {et~al.}(2020)\citenamefont
  {Yaouanc}, \citenamefont {{Dalmas de R{\'{e}}otier}}, \citenamefont
  {Roessli}, \citenamefont {Maisuradze}, \citenamefont {Amato}, \citenamefont
  {Andreica},\ and\ \citenamefont {Lapertot}}]{Yaouanc2020}%
  \BibitemOpen
  \bibfield  {author} {\bibinfo {author} {\bibfnamefont {A.}~\bibnamefont
  {Yaouanc}}, \bibinfo {author} {\bibfnamefont {P.}~\bibnamefont {{Dalmas de
  R{\'{e}}otier}}}, \bibinfo {author} {\bibfnamefont {B.}~\bibnamefont
  {Roessli}}, \bibinfo {author} {\bibfnamefont {A.}~\bibnamefont {Maisuradze}},
  \bibinfo {author} {\bibfnamefont {A.}~\bibnamefont {Amato}}, \bibinfo
  {author} {\bibfnamefont {D.}~\bibnamefont {Andreica}}, \ and\ \bibinfo
  {author} {\bibfnamefont {G.}~\bibnamefont {Lapertot}},\ }\href {\doibase
  10.1103/PhysRevResearch.2.013029} {\bibfield  {journal} {\bibinfo  {journal}
  {Phys. Rev. Res.}\ }\textbf {\bibinfo {volume} {2}},\ \bibinfo {pages}
  {013029} (\bibinfo {year} {2020})}\BibitemShut {NoStop}%
\bibitem [{\citenamefont {Bombor}\ \emph {et~al.}(2013)\citenamefont {Bombor},
  \citenamefont {Blum}, \citenamefont {Volkonskiy}, \citenamefont {Rodan},
  \citenamefont {Wurmehl}, \citenamefont {Hess},\ and\ \citenamefont
  {B{\"{u}}chner}}]{Bombor2013}%
  \BibitemOpen
  \bibfield  {author} {\bibinfo {author} {\bibfnamefont {D.}~\bibnamefont
  {Bombor}}, \bibinfo {author} {\bibfnamefont {C.~G.~F.}\ \bibnamefont {Blum}},
  \bibinfo {author} {\bibfnamefont {O.}~\bibnamefont {Volkonskiy}}, \bibinfo
  {author} {\bibfnamefont {S.}~\bibnamefont {Rodan}}, \bibinfo {author}
  {\bibfnamefont {S.}~\bibnamefont {Wurmehl}}, \bibinfo {author} {\bibfnamefont
  {C.}~\bibnamefont {Hess}}, \ and\ \bibinfo {author} {\bibfnamefont
  {B.}~\bibnamefont {B{\"{u}}chner}},\ }\href {\doibase
  10.1103/PhysRevLett.110.066601} {\bibfield  {journal} {\bibinfo  {journal}
  {Phys. Rev. Lett.}\ }\textbf {\bibinfo {volume} {110}},\ \bibinfo {pages}
  {066601} (\bibinfo {year} {2013})}\BibitemShut {NoStop}%
\bibitem [{\citenamefont {Kathyat}\ \emph {et~al.}(2020)\citenamefont
  {Kathyat}, \citenamefont {Mukherjee},\ and\ \citenamefont
  {Kumar}}]{Kathyat2020a}%
  \BibitemOpen
  \bibfield  {author} {\bibinfo {author} {\bibfnamefont {D.~S.}\ \bibnamefont
  {Kathyat}}, \bibinfo {author} {\bibfnamefont {A.}~\bibnamefont {Mukherjee}},
  \ and\ \bibinfo {author} {\bibfnamefont {S.}~\bibnamefont {Kumar}},\ }\href
  {\doibase 10.1103/PhysRevB.102.075106} {\bibfield  {journal} {\bibinfo
  {journal} {Phys. Rev. B}\ }\textbf {\bibinfo {volume} {102}},\ \bibinfo
  {pages} {075106} (\bibinfo {year} {2020})}\BibitemShut {NoStop}%
\bibitem [{\citenamefont {Banerjee}\ \emph {et~al.}(2014)\citenamefont
  {Banerjee}, \citenamefont {Rowland}, \citenamefont {Erten},\ and\
  \citenamefont {Randeria}}]{Banerjee2014}%
  \BibitemOpen
  \bibfield  {author} {\bibinfo {author} {\bibfnamefont {S.}~\bibnamefont
  {Banerjee}}, \bibinfo {author} {\bibfnamefont {J.}~\bibnamefont {Rowland}},
  \bibinfo {author} {\bibfnamefont {O.}~\bibnamefont {Erten}}, \ and\ \bibinfo
  {author} {\bibfnamefont {M.}~\bibnamefont {Randeria}},\ }\href {\doibase
  10.1103/PhysRevX.4.031045} {\bibfield  {journal} {\bibinfo  {journal} {Phys.
  Rev. X}\ }\textbf {\bibinfo {volume} {4}},\ \bibinfo {pages} {031045}
  (\bibinfo {year} {2014})}\BibitemShut {NoStop}%
\bibitem [{\citenamefont {Agarwala}\ and\ \citenamefont
  {Shenoy}(2017)}]{Agarwala2017}%
  \BibitemOpen
  \bibfield  {author} {\bibinfo {author} {\bibfnamefont {A.}~\bibnamefont
  {Agarwala}}\ and\ \bibinfo {author} {\bibfnamefont {V.~B.}\ \bibnamefont
  {Shenoy}},\ }\href {\doibase 10.1103/PhysRevLett.118.236402} {\bibfield
  {journal} {\bibinfo  {journal} {Phys. Rev. Lett.}\ }\textbf {\bibinfo
  {volume} {118}},\ \bibinfo {pages} {236402} (\bibinfo {year}
  {2017})}\BibitemShut {NoStop}%
\bibitem [{\citenamefont {Yang}\ \emph {et~al.}(2019)\citenamefont {Yang},
  \citenamefont {Qin}, \citenamefont {Deng}, \citenamefont {Duan},\ and\
  \citenamefont {Xu}}]{Yang2019}%
  \BibitemOpen
  \bibfield  {author} {\bibinfo {author} {\bibfnamefont {Y.-B.}\ \bibnamefont
  {Yang}}, \bibinfo {author} {\bibfnamefont {T.}~\bibnamefont {Qin}}, \bibinfo
  {author} {\bibfnamefont {D.-L.}\ \bibnamefont {Deng}}, \bibinfo {author}
  {\bibfnamefont {L.-M.}\ \bibnamefont {Duan}}, \ and\ \bibinfo {author}
  {\bibfnamefont {Y.}~\bibnamefont {Xu}},\ }\href {\doibase
  10.1103/PhysRevLett.123.076401} {\bibfield  {journal} {\bibinfo  {journal}
  {Phys. Rev. Lett.}\ }\textbf {\bibinfo {volume} {123}},\ \bibinfo {pages}
  {076401} (\bibinfo {year} {2019})}\BibitemShut {NoStop}%
\bibitem [{\citenamefont {Ritz}\ \emph {et~al.}(2013)\citenamefont {Ritz},
  \citenamefont {Halder}, \citenamefont {Wagner}, \citenamefont {Franz},
  \citenamefont {Bauer},\ and\ \citenamefont {Pfleiderer}}]{Ritz2013}%
  \BibitemOpen
  \bibfield  {author} {\bibinfo {author} {\bibfnamefont {R.}~\bibnamefont
  {Ritz}}, \bibinfo {author} {\bibfnamefont {M.}~\bibnamefont {Halder}},
  \bibinfo {author} {\bibfnamefont {M.}~\bibnamefont {Wagner}}, \bibinfo
  {author} {\bibfnamefont {C.}~\bibnamefont {Franz}}, \bibinfo {author}
  {\bibfnamefont {A.}~\bibnamefont {Bauer}}, \ and\ \bibinfo {author}
  {\bibfnamefont {C.}~\bibnamefont {Pfleiderer}},\ }\href {\doibase
  10.1038/nature12023} {\bibfield  {journal} {\bibinfo  {journal} {Nature}\
  }\textbf {\bibinfo {volume} {497}},\ \bibinfo {pages} {231} (\bibinfo {year}
  {2013})}\BibitemShut {NoStop}%
\bibitem [{\citenamefont {Yang}\ \emph {et~al.}()\citenamefont {Yang},
  \citenamefont {Chen}, \citenamefont {Yang},\ and\ \citenamefont
  {Dong}}]{Yang2005}%
  \BibitemOpen
  \bibfield  {author} {\bibinfo {author} {\bibfnamefont {H.-T.}\ \bibnamefont
  {Yang}}, \bibinfo {author} {\bibfnamefont {J.-W.}\ \bibnamefont {Chen}},
  \bibinfo {author} {\bibfnamefont {L.-F.}\ \bibnamefont {Yang}}, \ and\
  \bibinfo {author} {\bibfnamefont {J.}~\bibnamefont {Dong}},\ }\href {\doibase
  10.1103/PhysRevB.71.085402} {\ 10.1103/PhysRevB.71.085402}\BibitemShut
  {NoStop}%
\bibitem [{\citenamefont {Rastei}\ \emph {et~al.}(2007)\citenamefont {Rastei},
  \citenamefont {Heinrich}, \citenamefont {Limot}, \citenamefont {Ignatiev},
  \citenamefont {Stepanyuk}, \citenamefont {Bruno},\ and\ \citenamefont
  {Bucher}}]{Rastei2007}%
  \BibitemOpen
  \bibfield  {author} {\bibinfo {author} {\bibfnamefont {M.~V.}\ \bibnamefont
  {Rastei}}, \bibinfo {author} {\bibfnamefont {B.}~\bibnamefont {Heinrich}},
  \bibinfo {author} {\bibfnamefont {L.}~\bibnamefont {Limot}}, \bibinfo
  {author} {\bibfnamefont {P.~A.}\ \bibnamefont {Ignatiev}}, \bibinfo {author}
  {\bibfnamefont {V.~S.}\ \bibnamefont {Stepanyuk}}, \bibinfo {author}
  {\bibfnamefont {P.}~\bibnamefont {Bruno}}, \ and\ \bibinfo {author}
  {\bibfnamefont {J.~P.}\ \bibnamefont {Bucher}},\ }\href {\doibase
  10.1103/PhysRevLett.99.246102} {\bibfield  {journal} {\bibinfo  {journal}
  {Phys. Rev. Lett.}\ }\textbf {\bibinfo {volume} {99}},\ \bibinfo {pages}
  {246102} (\bibinfo {year} {2007})}\BibitemShut {NoStop}%
\bibitem [{\citenamefont {Tejo}\ \emph {et~al.}(2018)\citenamefont {Tejo},
  \citenamefont {Riveros}, \citenamefont {Escrig}, \citenamefont {Guslienko},\
  and\ \citenamefont {Chubykalo-Fesenko}}]{Tejo2018a}%
  \BibitemOpen
  \bibfield  {author} {\bibinfo {author} {\bibfnamefont {F.}~\bibnamefont
  {Tejo}}, \bibinfo {author} {\bibfnamefont {A.}~\bibnamefont {Riveros}},
  \bibinfo {author} {\bibfnamefont {J.}~\bibnamefont {Escrig}}, \bibinfo
  {author} {\bibfnamefont {K.~Y.}\ \bibnamefont {Guslienko}}, \ and\ \bibinfo
  {author} {\bibfnamefont {O.}~\bibnamefont {Chubykalo-Fesenko}},\ }\href
  {\doibase 10.1038/s41598-018-24582-x} {\bibfield  {journal} {\bibinfo
  {journal} {Sci. Rep.}\ }\textbf {\bibinfo {volume} {8}},\ \bibinfo {pages}
  {6280} (\bibinfo {year} {2018})}\BibitemShut {NoStop}%
\bibitem [{\citenamefont {Kumar}\ and\ \citenamefont
  {Majumdar}(2006)}]{Kumar2006c}%
  \BibitemOpen
  \bibfield  {author} {\bibinfo {author} {\bibfnamefont {S.}~\bibnamefont
  {Kumar}}\ and\ \bibinfo {author} {\bibfnamefont {P.}~\bibnamefont
  {Majumdar}},\ }\href {\doibase 10.1140/epjb/e2006-00173-2} {\bibfield
  {journal} {\bibinfo  {journal} {Eur. Phys. J. B}\ }\textbf {\bibinfo {volume}
  {50}},\ \bibinfo {pages} {571} (\bibinfo {year} {2006})}\BibitemShut
  {NoStop}%
\bibitem [{\citenamefont {Mukherjee}\ \emph {et~al.}(2015)\citenamefont
  {Mukherjee}, \citenamefont {Patel}, \citenamefont {Bishop},\ and\
  \citenamefont {Dagotto}}]{Mukherjee2015a}%
  \BibitemOpen
  \bibfield  {author} {\bibinfo {author} {\bibfnamefont {A.}~\bibnamefont
  {Mukherjee}}, \bibinfo {author} {\bibfnamefont {N.~D.}\ \bibnamefont
  {Patel}}, \bibinfo {author} {\bibfnamefont {C.}~\bibnamefont {Bishop}}, \
  and\ \bibinfo {author} {\bibfnamefont {E.}~\bibnamefont {Dagotto}},\ }\href
  {\doibase 10.1103/PhysRevE.91.063303} {\bibfield  {journal} {\bibinfo
  {journal} {Phys. Rev. E}\ }\textbf {\bibinfo {volume} {91}},\ \bibinfo
  {pages} {063303} (\bibinfo {year} {2015})}\BibitemShut {NoStop}%
\bibitem [{\citenamefont {Loring}\ and\ \citenamefont
  {Hastings}(2010)}]{Loring2010}%
  \BibitemOpen
  \bibfield  {author} {\bibinfo {author} {\bibfnamefont {T.~A.}\ \bibnamefont
  {Loring}}\ and\ \bibinfo {author} {\bibfnamefont {M.~B.}\ \bibnamefont
  {Hastings}},\ }\href {\doibase 10.1209/0295-5075/92/67004} {\bibfield
  {journal} {\bibinfo  {journal} {EPL (Europhysics Lett.}\ }\textbf {\bibinfo
  {volume} {92}},\ \bibinfo {pages} {67004} (\bibinfo {year}
  {2010})}\BibitemShut {NoStop}%
\bibitem [{\citenamefont {Huang}\ and\ \citenamefont
  {Liu}(2018{\natexlab{a}})}]{Huang2018}%
  \BibitemOpen
  \bibfield  {author} {\bibinfo {author} {\bibfnamefont {H.}~\bibnamefont
  {Huang}}\ and\ \bibinfo {author} {\bibfnamefont {F.}~\bibnamefont {Liu}},\
  }\href {\doibase 10.1103/PhysRevLett.121.126401} {\bibfield  {journal}
  {\bibinfo  {journal} {Phys. Rev. Lett.}\ }\textbf {\bibinfo {volume} {121}},\
  \bibinfo {pages} {126401} (\bibinfo {year} {2018}{\natexlab{a}})}\BibitemShut
  {NoStop}%
\bibitem [{\citenamefont {Huang}\ and\ \citenamefont
  {Liu}(2018{\natexlab{b}})}]{Huang2018a}%
  \BibitemOpen
  \bibfield  {author} {\bibinfo {author} {\bibfnamefont {H.}~\bibnamefont
  {Huang}}\ and\ \bibinfo {author} {\bibfnamefont {F.}~\bibnamefont {Liu}},\
  }\href {\doibase 10.1103/PhysRevB.98.125130} {\bibfield  {journal} {\bibinfo
  {journal} {Phys. Rev. B}\ }\textbf {\bibinfo {volume} {98}},\ \bibinfo
  {pages} {125130} (\bibinfo {year} {2018}{\natexlab{b}})}\BibitemShut
  {NoStop}%
\end{thebibliography}

%

\end{document}